\documentclass[acmsmall,screen,nonacm]{acmart}

\usepackage{booktabs}      
\usepackage{listings}      
\usepackage{xcolor}
\usepackage{graphicx}
\usepackage{rotating} 
\usepackage{algorithm2e}
\usepackage{tikz}
\usetikzlibrary{arrows.meta,positioning,calc}
\usepackage{makecell}
\usepackage{booktabs}
\usepackage{caption}
\usepackage{ragged2e}
\usepackage{multirow}
\usepackage{array} 
\usepackage{cleveref}
\usepackage{placeins}
\usepackage[frozencache,cachedir=.]{minted}
\usepackage{subcaption}
\usepackage[most]{tcolorbox}
\usepackage{wrapfig}
\usepackage{tcolorbox}
\usepackage{etoolbox}
\tcbuselibrary{breakable}
\usepackage{placeins} 

\usepackage{enumitem}
\setlist{leftmargin=7mm}
\usepackage{xcolor}

\newif\ifrevision
\revisionfalse 
\ifrevision
  \newcommand{\rev}[1]{\textcolor{red}{#1}}
\else
  \newcommand{\rev}[1]{#1}
\fi

\ifrevision
  
\else
  
\fi
\usepackage{xcolor}

\ifrevision
  \newenvironment{revblock}{\color{red}}{}
\else
  \newenvironment{revblock}{}{}
\fi
\newtcbox{\hly}{on line,
    arc=0pt,
    colframe=yellow!50, 
    colback=yellow!50, 
    boxrule=0.0pt, 
    boxsep=0pt, 
    left=1pt, right=1pt, top=2pt, bottom=2pt 
}

\newtcbox{\hlgr}{on line,
    arc=0pt,
    colframe=gray!30, 
    colback=gray!30, 
    boxrule=0.0pt, 
    boxsep=0pt, 
    left=1pt, right=1pt, top=2pt, bottom=2pt 
}

\newtcbox{\hlg}{on line,
    arc=0pt,
    colframe=green!50, 
    colback=green!50, 
    boxrule=0.0pt, 
    boxsep=0pt, 
    left=1pt, right=1pt, top=2pt, bottom=2pt 
}

\tcbset{
    mybox/.style={colback=white, width=\textwidth, boxrule=0.2mm,title=Prompt, sharp corners,left=1mm, right=1mm, bottom=1mm},
}
\tcbset{enhanced,attach boxed title to top left={xshift=2mm,yshift=-2mm},coltitle=black,boxed title style={size=small,colback=white, boxrule=0.2mm}}

\newcommand{\projName}{\textsc{Gordian}\xspace}
\lstdefinelanguage{text}{
  moredelim=[is][\ttfamily]{\%\%}{\%\%},
  basicstyle=\ttfamily,
}

\lstdefinestyle{CStyle}{
  language=C,
  basicstyle=\ttfamily\footnotesize,
  keywordstyle=\color{blue},
  commentstyle=\color{gray},
  stringstyle=\color{orange!60!black},
  numbers=left,
  numberstyle=\tiny\color{gray},
  stepnumber=1,
  frame=single,
  xleftmargin=10pt,
  xrightmargin=10pt,
  breaklines=true,
  tabsize=2
}

\newcommand{\codelist}[2]{%
  \begin{lstlisting}[style=CStyle,caption={#1}]
#2
  \end{lstlisting}
}

\theoremstyle{definition}

\title[Defusing Logic Bombs in Symbolic Execution with LLM-Generated Ghost Code]{Defusing Logic Bombs in Symbolic Execution\\ with LLM-Generated Ghost Code}

\author{Dimitrios Stamatios Bouras}
\email{2501112125@stu.pku.edu.cn}
\affiliation{%
  \institution{Key Lab of HCST (PKU), MoE, SCS, Peking University}
  \city{Beijing}
  \country{China}
}

\author{Sergey Mechtaev}
\authornote{Sergey Mechtaev is the corresponding author.}
\email{mechtaev@pku.edu.cn}
\affiliation{%
  \institution{Key Lab of HCST (PKU), MoE, SCS, Peking University}
  \city{Beijing}
  \country{China}}

\begin{document}

\begin{abstract}
    Symbolic execution is a powerful program analysis technique, but its effectiveness is fundamentally limited by solver-hostile program fragments, complex numerical reasoning, and unbounded heap structures. Recent work proposed replacing constraint solvers with large language models (LLMs) to bypass these limitations, but such approaches struggle to analyze real-world codebases, where deep execution paths require globally consistent reasoning across many interacting constraints. We present \projName, a hybrid symbolic execution framework that uses LLMs selectively to generate lightweight ghost code that aids an SMT solver in handling solver-hostile code fragments, while preserving its precise, global reasoning capability. In particular, we propose three types of ghost code: (1) inversion of difficult code fragments with iterative bidirectional constraint propagation, (2) modeling via solver-friendly surrogates while preserving relevant behavior, and (3) semantic partitioning of unbounded heap spaces. We implemented \projName on top of the KLEE symbolic execution engine and evaluated it on synthetic ``logic bombs'' capturing distinct symbolic reasoning challenges, a popular mathematical library FDLibM, and \rev{four} structured-input programs (\texttt{libexpat}, \texttt{jq}, \texttt{bc} and \texttt{libyaml}). Across benchmarks, \projName{} improves coverage by \rev{28.5--115.2\%  over traditional symbolic execution baseline and by 74.1--189.8\% over LLM-based symbolic execution baselines}, while reducing LLM token usage by an average of \rev{91--96\%}. This highlights the practicality and effectiveness of this approach in real-world settings.
\end{abstract}

\maketitle

\section{Introduction}

Symbolic execution runs code on abstract inputs, encodes program behavior in logical constraints, and uses an SMT solver to analyze this behavior. The applicability of symbolic execution is hindered by several inherent limitations. First, because SMT solving is theoretically undecidable in general, symbolic execution often struggles with programs involving complex computations, as these translate into correspondingly complex SMT constraints. Second, symbolic execution requires precise modeling of program behavior, which becomes challenging in real-world environments, such as those involving network communication or extensive system calls, that are difficult to model accurately. Third, the number of possible program execution paths is effectively infinite, meaning symbolic execution can only explore a limited subset. The above challenges are particularly acute in programs that make extensive use of dynamic memory allocation and pointer-based data structures, where control flow depends on complex aliasing relations or heap configurations.

Recent developments in large language models (LLMs) promise to mitigate these limitations. In particular, Autobug~\cite{autobug}, PALM~\cite{palm} and ConcoLLMic~\cite{concollmic} use LLMs in place of, or to substantially reduce reliance on, SMT solving to directly analyze the program and generate tests for difficult program parts. While this circumvents the undecidability issues of SMT solvers, these methods struggle to analyze real-world codebases when the target functionality is deep in the program,  reachable only after several layers of calls and gating checks that require globally consistent choices about inputs and state. This is because LLMs struggle to maintain long-range, formally checkable dependencies across many functions, whereas SMT solvers accumulate and satisfy precise cross-procedural constraints, which enables reasoning about subtle interprocedural interactions.

We propose \projName, a novel way of integrating symbolic execution with LLMs. Instead of replacing SMT solvers, \projName uses an LLM only where SMT-based symbolic execution struggles. In particular, an LLM analyzes difficult parts of the program, recognizing domain structures and identifying constraint solving strategies, and translates these into lightweight ``ghost code'' that plugs into the SMT-based symbolic execution workflow as domain-specific solver aids. This preserves precise, global reasoning over cross-function dependencies and subtle interactions, while alleviating hard constraints that block progress in traditional symbolic execution.

We introduce three types of LLM-generated ghost code for helping symbolic execution analyze a complex program fragment. First, we prompt an LLM to invert the difficult fragment, so that it can be executed backward to propagate concrete values from the suffix constraints back across the fragment to satisfy prefix constraints. Second, we prompt an LLM to generate a simplified surrogate that approximates complex behavior to permit efficient forward symbolic execution. Third, for program fragments operating over dynamic data structures, we prompt an LLM to generate symbolic heap topologies based on domain-specific insights to mitigate path explosion. The choice of ghost code type is context-dependent and selected by the LLM. Soundness is preserved because any inputs or states produced with ghost code are checked against the original program, so hallucinated ghost code does not introduce unsound results.

We implemented \projName in the KLEE symbolic execution engine~\cite{klee} and compared it against previous techniques on programs that have historically challenged symbolic execution: (1) 53 synthetic programs from the LogicBombs benchmark~\cite{xu2018benchmarking} capturing distinct symbolic-reasoning challenges; (2) 78 entry points\footnote{Entry points are top-level target functions that exercise different parts of the program or library.} from the widely-used math library \texttt{FDLibM}, which induces constraints involving undecidable theories; and (3) \rev{19} source files from \rev{four} projects (\texttt{libexpat}, \texttt{jq}, \texttt{bc} and \texttt{libyaml}) targeting structured input formats and thus operating on complex dynamic data structures. Across the benchmarks, \projName{} improves coverage by \rev{28.5--115.2\% over the strongest traditional symbolic-execution baseline
and by 74.1--189.8\% over the strongest relevant LLM-based baseline} while reducing the number of tokens by \rev{91--96\%}. Via ablation, we confirmed that each type of ghost code contributed to the performance improvement.

In summary, the paper makes the following contributions:

\begin{itemize}
\item A novel enhancement of symbolic execution with LLM-generated solver-aiding ghost code.
\item Three practical ghost code variants, as well as corresponding generation prompts and solver integration algorithms, to alleviate practical challenges of symbolic execution.
\item A comprehensive evaluation on both synthetic and realistic benchmarks that highlights the strength of our approach in covering hard-to-reach program branches.
\end{itemize}

All code, data, and scripts are available at \url{https://figshare.com/s/09550df85983cbf4ae71}.

\section{Overview}

Ghost code~\cite{filliatre2016spirit} refers to auxiliary constructs added to a program to facilitate specification, reasoning, or verification without affecting its functional behavior. Traditionally, ghost code has been used in program verification to express invariants, ranking functions, or lemmas to automate proofs of program properties. Our work is the first that applies ghost code to enhance symbolic execution, and shows that it can be automatically generated using LLMs. This section illustrates the utility of LLM-generated ghost code for symbolically executing programs that involve non-trivial mathematics and dynamic data structures.

\begin{figure}
  \centering
  \begin{subfigure}[b]{0.24\textwidth}
    \centering
{
\scriptsize          
\begin{minted}[escapeinside=||]{C}
hx = __HI(x);
ix = hx & 0x7fffffff;
if (ix >= 0x7ff00000)
  return one / (x * x);
x = fabs(x);
if (ix >= 0x40000000) {
  |\hly{s = sin(x);}|
  |\hly{c = cos(x);}|
  ss = s - c;
  cc = s + c;
  if (ix < 0x7fe00000) {
    z = -cos(x + x);
    if (|\hlg{(s * c) < 0}|)
      cc = z / ss;
    else
      ss = z / cc;
    }
}
\end{minted}
}
    \caption{Function \texttt{\_ieee754\_j0} from \texttt{FDLibM} involving trigonometric operations.}
    \label{fig:ieee754}
  \end{subfigure}
  \hfill
  \begin{subfigure}[b]{0.36\textwidth}
    \centering
{
\tiny          
\begin{minted}[escapeinside=||]{C}
double golden_section(double (*f)(double, void *),
                      void *ctx,
                      double a, double b,
                      int max_iter,
                      double tol) {
  |\hlgr{algorithm omitted for brevity}|
}
                      
double sincos_error(double x, void *ctx) {
  double s = sin(x), c = cos(x);
  double ds = s - ((double *)ctx)[0];
  double dc = c - ((double *)ctx)[1];
  return ds * ds + dc * dc;
}

double f_inv(double s, double c) {
  double ctx[2] = {s, c};
  return golden_section(
    sincos_error, ctx, 0.0, 2*M_PI, 200, 1e-12
  );
}
\end{minted}
}
    \caption{LLM-generated ghost code that inverts \mintinline{C}{s = sin(x); c = cos(x);} using golden-section search algorithm.}
    \label{fig:inversion}
  \end{subfigure}
  \hfill
  \begin{subfigure}[b]{0.35\textwidth}
    \centering 
    {\relscale{0.7}\def\svgwidth{1.4\textwidth}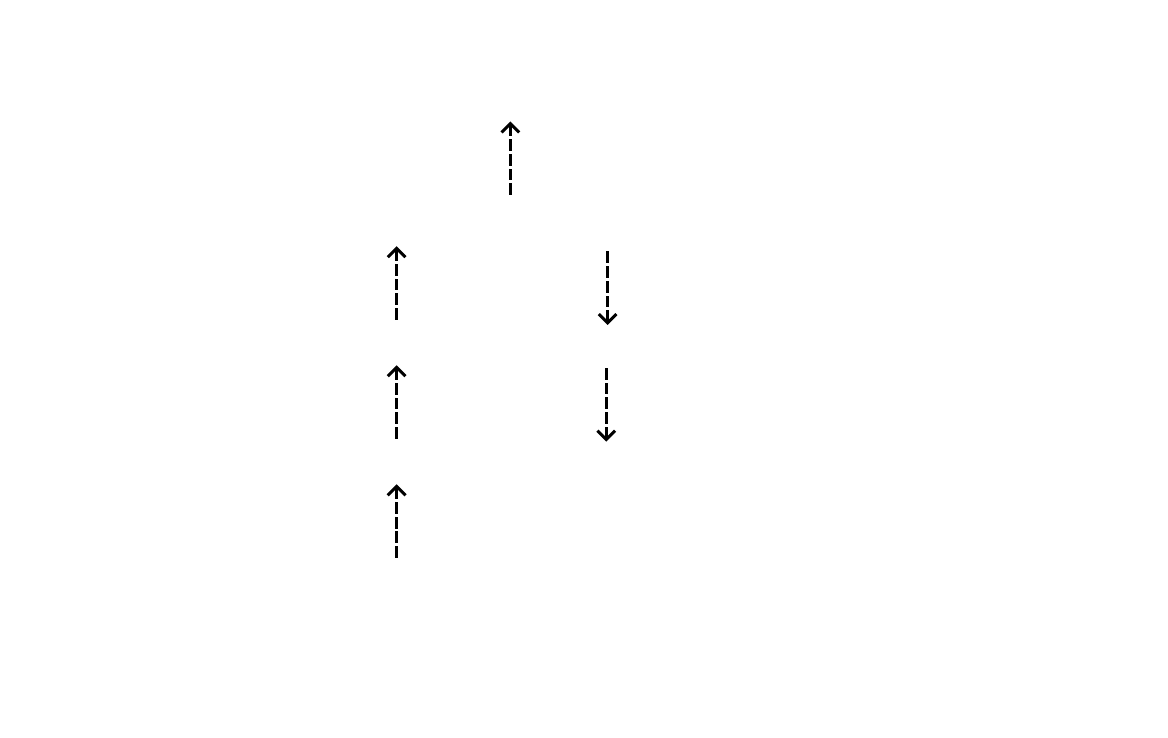}\vspace{2mm}
    \caption{sin/cos are replaced with $\alpha, \beta$ during symbolic execution; values are propagated backward by running $f^{-1}$.}
    \label{fig:propagation}
  \end{subfigure}
  \vspace{-2mm}
  \caption{\projName propagates constraints through a difficult code fragment that involves trigonometric operations using a concrete execution of an LLM-generated ghost function that inverts the fragment.}
  \label{fig:complex_math_example}
  \vspace{-2mm}
\end{figure}

\subsection{Propagating Constraints Through Difficult Code Fragments}

Symbolic execution struggles to analyze programs with complex operations such as non-linear arithmetic or trigonometric functions, since they induce hard-to-solve constraints. Consider an implementation of the Bessel function in \Cref{fig:ieee754}, which is a part of the function \texttt{\_ieee754\_j0} from the math library \texttt{FDLibM}. Suppose symbolic execution aims to explore branches guarded by the condition \mintinline[escapeinside=||]{C}`(s * c) < 0` highlighted with \hlg{\phantom{X}}. Even KLEE-Float~\cite{kleeFloat}, a version of KLEE~\cite{klee} optimized for floating-point operations, fails to solve the resulting constraints, because this check depends on the output of trigonometric operations highlighted with \hly{\phantom{X}}.

As an alternative to traditional techniques, ConcoLLMic~\cite{concollmic} replaces the SMT solver with an agentic LLM workflow to generate inputs that cover target branches. It formulates queries of the form $f(a) = b$ and then  synthesizes a suitable input $a$ using natural-language reasoning and Python tool execution. In \Cref{fig:ieee754}, although ConcoLLMic can find a satisfying assignment for constraints originating from the difficult fragment highlighted by \hly{\phantom{X}}, this fragment itself depends on the value of \mintinline[escapeinside=||]{C}`x`. The variable \mintinline[escapeinside=||]{C}`x` induces a non-trivial control-flow dependency through bit-level manipulation of its IEEE-754 representation: (1) extracting the high 32 bits with \mintinline{C}`__HI(x)`, (2) masking off the sign bit with \mintinline{C}`hx & 0x7fffffff` to obtain $|x|$, and (3) checking the guard \mintinline{C}`ix >= 0x40000000`, which corresponds to $|x| \ge 2.0$. Because ConcoLLMic lacks a mechanism for global reasoning that can combine constraints arising from multiple program locations, it fails to infer and satisfy these intertwined conditions, and therefore cannot cover the target branches guarded by \mintinline[escapeinside=||]{C}`(s * c) < 0`.

\begin{figure}
\vspace{-3mm}
  \centering
  \begin{subfigure}[b]{0.32\textwidth}
    \centering
{
\scriptsize          
\begin{minted}[escapeinside=@@]{C}
typedef struct SNode {
    int val;
    char *name;
    struct SNode *lvl0;
    struct SNode *lvl1;
} SNode;
  
int proc_if_ring(SNode *h, SNode *g) {
  SNode *a=h ? h->lvl0 : NULL;
  SNode *b=a ? a->lvl0 : NULL;
  SNode *c=b ? b->lvl0 : NULL;
  SNode *d=c ? c->lvl0 : NULL;
  if (!h||!a||!b||!c||!d) return 0;
  if (d!=h) return 0;
  if (h->lvl1!=b||a->lvl1!=c||
      b->lvl1!=d||c->lvl1!=a) return 0;
  SNode *g1 = g ? g->lvl0 : NULL;
  SNode *g2 = g1 ? g1->lvl0 : NULL;
  if (g2!=h) return 0;
  @\hlg{return proc\_ring\_values(h)}@
}
\end{minted}
}
    \caption{The function \texttt{proc\_if\_ring} checks if a skiplist forms a ring, and then processes its values.}
    \label{fig:process_ring}
  \end{subfigure}
  \hfill
  \begin{subfigure}[b]{0.28\textwidth}
    \centering
\hspace*{-2mm}\resizebox{1.2\linewidth}{!}{%
\begin{tikzpicture}[
    >=stealth,
    node distance=2cm,
    main/.style={draw,rounded corners,minimum width=11mm,minimum height=8mm,align=center},
    arr/.style={->,very thick},
    exp/.style={->,densely dashed,thick,black!70},
    gptr/.style={->,ultra thick,red!70!black,shorten >=1pt,shorten <=2pt},
    x=1cm,y=1cm
  ]
  \centering
  \node[main] (n0) at (-1, 1.5) {n0};
  \node[main] (n1) at ( 1, 1.5) {n1};
  \node[main] (n2) at ( 1,-1.5) {n2};
  \node[main] (n3) at (-1,-1.5) {n3};

  \draw[arr] (n0) -- node[draw=none,fill=none,above]{lvl0} (n1);
  \draw[arr] (n1) -- node[draw=none,fill=none,right]{lvl0} (n2);
  \draw[arr] (n2) -- node[draw=none,fill=none,below]{lvl0} (n3);
  \draw[arr] (n3) -- node[draw=none,fill=none,left]{lvl0} (n0);

    \node[draw=none,fill=none,above left=15pt and 7pt of n0] (glabel) {$h$};
  \draw[gptr] (glabel.north east) -- (n0.north);

  \draw[exp]
    (n0) .. controls ($(n0)+(0,3.8)$) and ($(n2)+(0,3.8)$) ..
    node[draw=none,fill=none,above]{lvl1} (n2.north west);
  \draw[exp]
    (n1) .. controls ($(n1)+(2.8,0)$) and ($(n3)+(2.8,0)$) ..
    node[draw=none,fill=none,right]{lvl1} (n3.north east);
  \draw[exp]
    (n2) .. controls ($(n2)+(0,-3.8)$) and ($(n0)+(0,-3.8)$) ..
    node[draw=none,fill=none,below]{lvl1} (n0.south east);
\draw[exp]
    (n3) .. controls ($(n3)+(-2.8,0)$) and ($(n1)+(-2.8,0)$) ..
    node[draw=none,fill=none,left]{lvl1} (n1.south west);

  \node[draw=none,fill=none,below right=15pt and 7pt of n2] (glabel) {$g$};
  \draw[gptr] (glabel.north west) -- (n2.south);
\end{tikzpicture}
}
    \caption{Ring-shape skiplist required to reach the target function call \texttt{proc\_ring\_values}.}
    \label{fig:ring}
  \end{subfigure}
  \hfill
  \begin{subfigure}[b]{0.36\textwidth}
    \centering
{
\tiny          
\begin{minted}[escapeinside=!!]{C}
void constrain_heap(SNode *n0, SNode *n1,
                    SNode *n2, SNode *n3,
                    SNode **h, SNode **g) {
  switch (!$\xi$!) {
    case 1: /* broken chain */
      n0->lvl0 = n1; n1->lvl0 = NULL;
      n0->lvl1 = NULL; n1->lvl1 = NULL;
      n2->lvl1 = NULL; n3->lvl1 = NULL;
      *h = n0; *g = n3;
      break;

    !\hlgr{three cases omitted for brevity}!

    case 5: /* ring + g two-steps to h */
      n0->lvl0 = n1; n1->lvl0 = n2;
      n2->lvl0 = n3; n3->lvl0 = n0;
      n0->lvl1 = n2; n1->lvl1 = n3;
      n2->lvl1 = n0; n3->lvl1 = n1;
      *h = n0;
      *g = n2;
      break;
  }
}
\end{minted}
}
    \caption{LLM-generated ghost code that constrains heap with five semantic topologies relevant to the problem context.}
    \label{fig:constrain_heap}
  \end{subfigure}
  \vspace{-2mm}
  \caption{\projName tackles heap-configuration explosion by using LLM-generated ghost code to constrain the heap to semantically meaningful data-structure shapes while preserving symbolic contents.}
  \label{fig:topology_example}
  \vspace{-2mm}  
\end{figure}

To address this challenge, \projName operates in three steps. In step 1, it asks an LLM to identify a code fragment that is hard for an SMT solver to reason about and to generate ghost code that inverts the fragment. In this example, the LLM flags \mintinline{C}`s = sin(x); c = cos(x);`, which we refer to as $f$, as difficult and produces an efficient inverse procedure \(f^{-1}\) using golden-section search~\cite{kiefer1953sequential} as shown in \Cref{fig:inversion}. In step 2, \projName executes the program on a symbolic input \(x\), replacing the difficult fragment with fresh symbolic variables \(\alpha\) and \(\beta\) that stand for the values of \(s\) and \(c\) as shown in \Cref{fig:propagation}. During this run it collects two constraints: (1) a prefix path condition \(\pi_{\mathrm{pre}}\) from code preceding the fragment, which depends only on the input \(x\); and (2) a suffix path condition \(\pi_{\mathrm{suf}}\) from code following the fragment, which depends on \(\alpha\) and \(\beta\).

The final step integrates the synthesized ghost code into the constraint-solving workflow via \emph{bidirectional constraint propagation}. First, \projName solves the suffix constraints \(\pi_{\mathrm{suf}}\) with an SMT solver to obtain candidate values for \(s\) and \(c\), denoted \(\alpha_0\) and \(\beta_0\); assume the solver returns \((\alpha_0,\beta_0)=(10,-1)\). Next, \projName propagates these values backward through the difficult fragment by concretely executing the synthesized inverse \(f^{-1}\). Although no real \(x\) can satisfy these values simultaneously (since \(\sin^2(x)+\cos^2(x)=1\)), the inverse computation yields the closest feasible input \(x_0 \approx 1.67\) radians. This input violates the prefix constraint \(\pi_{\mathrm{pre}}\), which requires \(|x|\ge 2.0\). \projName therefore uses an optimization solver to find the closest \(x_1\) to \(x_0\) that satisfies \(\pi_{\mathrm{pre}}\); in this example, \(x_1=2.0\). It then propagates \(x_1\) forward by executing the original fragment \(f\), producing \((\alpha_1,\beta_1)\approx(0.909,-0.416)\), which satisfies \(\pi_{\mathrm{suf}}\). Finally, \projName concretely executes the program to validate that the generated input indeed reaches the target branch, mitigating potential ghost code inaccuracies due to LLM hallucinations.

In comparison to traditional symbolic execution, \projName takes advantage of LLM's domain knowledge and code generation capabilities in the form of an inversion algorithm, for which it chose to implement a golden-section search. In comparison to LLM-based techniques such as ConcoLLMic, \projName takes advantage of the SMT solver's precise reasoning about cross-function interactions, i.e. \mintinline{C}{_ieee754_j0} and \mintinline{C}{__HI}, achieved through our bidirectional constraint propagation algorithm that plugs the generated ghost code into the solver logic. Through this synergy, \projName is the only tool that provides full coverage for this function.

\subsection{Constraining Heap With Semantic Topologies}

A major limitation of symbolic execution in real-world testing is heap-configuration explosion caused by dynamic data structures, such as structured-input parse trees and the complex list/tree structures used in databases and file systems. Due to the complexity of these data structures and their processing code, we demonstrate how \projName addresses this problem on a synthetic example of a simplified 2-level skiplist shown in \Cref{fig:process_ring}. In this data structure, \mintinline{C}`lvl0` points to the next node in the full (level-0) linked list, while \mintinline{C}`lvl1` is an ``express lane'' pointer that skips over some nodes to speed up searches. Each node stores a key/value \mintinline{C}`val` and associated data \mintinline{C}`name`.

Let the function under test be \mintinline{C}{proc_if_ring} in \Cref{fig:process_ring}. It checks whether a given skiplist forms a valid ring, as in \Cref{fig:ring}, and, if so, processes the values stored in its nodes. Testing this function is challenging because reaching the call to the value-processing routine (highlighted with \hlg{\phantom{X}}) requires symbolic execution to explore a prohibitive number of heap configurations, covering many possible shapes and aliasing relationships among ten pointers. As a result, KLEE fails to reach this function within the timeout of 10 minutes. On the other hand, the LLM-based tool Cottontail~\cite{cottontail} accounts for input structure and can generate valid data-structure shapes, but --- similarly to ConcoLLMic --- it cannot reconcile value constraints originating elsewhere (e.g., in \texttt{proc\_ring\_values}), limiting its effectiveness on large projects.

To address these limitations, \projName leverages an LLM's contextual understanding to generate ghost code, \texttt{constrain\_heap}, which constrains the heap with \emph{semantic topologies}. A semantic topology is a partially concrete, partially symbolic heap configuration that captures a meaningful use case drawn from the problem domain or program context. In this example, the LLM generates five semantic topologies --- chain, non-ring, misaligned express pointers, mismatched head link, and a valid ring -- shown in \Cref{fig:constrain_heap}. The selected topology is determined by a fresh symbolic variable $\xi$. When symbolically executing the program, \projName inserts a call of \texttt{constrain\_heap} in front of each call of \mintinline{C}{proc_if_ring}:

{\scriptsize
\begin{minted}[escapeinside=@@]{C}
/* n0, n1, n2, n3, h, g are initialized with symbolic memory */
constrain_heap(n0, n1, n2, n3, h, g);
int result = proc_if_ring(h, g);
\end{minted}
}

\noindent Crucially, within each topology, only the links relevant to the intended shape are fixed; all other fields remain symbolic, including the value and name fields. This enables exploration of shape-dependent branches without over-constraining subsequent execution. As a result, \projName efficiently covers all branches in the function under test, as well as the function \texttt{proc\_ring\_values}, within 2 seconds.

\section{Background \& Notation}

We work in the standard Satisfiability Modulo Theories (SMT) setting. Let $X$ denote a finite set of (typed) variables, ranged over by $\alpha,\beta,\gamma,\xi\in X$, and let there be a fixed signature (function and predicate symbols) interpreted in a background theory $T$ (e.g., linear arithmetic).
We write $\Phi_{T}(X)$, or just $\Phi$ for short, for the set of all well-formed formulas with free variables in $X$; formulas are denoted by $\phi,\psi\in \Phi$.
An assignment (or model candidate) is a map $A: X \to \mathrm{Dom}(T)$; satisfaction is written $A \models_T \phi$.
A formula $\phi$ is $T$-satisfiable if there exists $A$ such that $A \models_T \phi$, and $T$-unsatisfiable otherwise.
We denote by $\mathrm{SMT}(\phi)$ an SMT-solver call that returns either \textsc{sat} together with a satisfying assignment/model $A$, or \textsc{unsat}.
For optimization, let $f$ be a (theory) term denoting an objective value (typically numeric). We write $\mathrm{OPT}(f,\phi)$ for an optimization-modulo-theories (OMT) call returning an assignment $A^\star$ such that $A^\star \models_T \phi$ and $f(A^\star)$ is optimal among all $A$ satisfying $\phi$ (with \textsc{min} or \textsc{max} understood from context), or returning \textsc{unsat} if $\phi$ is unsatisfiable.
For substitution, given a term $t$ and variable $x\in X$, we write $\phi[x\mapsto t]$ for the capture-avoiding substitution of $t$ for all free occurrences of $x$ in $\phi$.

Let $\mathcal{P}$ denote the set of programs, ranged over by $P,Q\in\mathcal{P}$, and let $\mathcal{S}$ denote the set of statements, ranged over by $s,t\in\mathcal{S}$; programs are (finite) sequences of statements, written $P = s_1;\cdots;s_n$. We assume a set of program variables $\mathsf{Var}$, denoted by $v,w$, distinct from the set of logical/symbolic variables $X$ used in formulas/constraints. Procedure calls have the form $\mathsf{call}\;p(e_1, ..., e_n)$ for $p\in\mathsf{Proc}$ and argument expressions $e_1, ..., e_n$; we write $P[C \mapsto \mathsf{call}\;p(e_1, ..., e_n)]$ for replacing a designated code fragment (subprogram) $C$ occurring in $P$ by that call.

A (concrete) program state is a pair $\sigma=(\nu,h)$ where the store $\nu:\mathsf{Var}\to\mathrm{Val}$ maps program variables to concrete values and the heap $h:\mathrm{Addr}\rightarrow \mathrm{Cell}$ is a finite partial map from addresses to heap cells (records), with $\mathrm{Cell}$ determined by the language (e.g., tuples of fields); a symbolic state is a triple $\hat\sigma=(\hat\nu,\hat h,\pi)$ where the symbolic store $\hat\nu:\mathsf{Var}\to\mathrm{Term}(X)$ maps variables to SMT terms, the symbolic heap $\hat h:\mathrm{Addr}\rightarrow \mathrm{CellTerm}(X)$ maps addresses to (possibly symbolic) cells whose fields are SMT terms, and $\pi\in\Phi_{T}(X)$ is the path condition constraining the symbols. All concrete states are denoted as $\Sigma$, and symbolic states as $\widehat{\Sigma}$. Satisfaction/realization is given by an assignment $A$ such that $A\models_T \pi$, yielding a concrete state $\sigma_A=(\nu_A,h_A)$ by evaluating all terms in $\hat\nu$ and $\hat h$ under $A$ (writing $\llbracket t\rrbracket_A$ for term evaluation). A concrete program semantics is a transition relation $\to \;\subseteq\; (\mathcal S\times \Sigma)\times(\mathcal S\times \Sigma)$, and its reflexive–transitive closure is written $\to^{*}$. Then, a concrete execution function is defined as $\mathrm{exec}(P,\sigma)\;=\;\sigma' \in \Sigma \text{ such that } \langle P,\sigma\rangle \to^{*} \langle \_,\sigma'\rangle.$

\begin{example}[Concrete execution of a simple assignment]
Let $P$ be the program $x := x + 1$ and let the initial store be
$\sigma$ with $\sigma(x)=41$. Since $\llbracket x+1\rrbracket_\sigma = 42$, we have $\langle x := x + 1,\sigma\rangle \to \langle \_, \sigma[x \mapsto 42]\rangle$. Therefore, $\mathrm{exec}(x := x+1,\sigma)=\sigma[x \mapsto 42]$.
\end{example}

A symbolic semantics is a transition relation $\Rightarrow \;\subseteq\; (\mathcal S\times \widehat\Sigma)\times(\mathcal S\times \widehat\Sigma)$ that updates symbolic store/heap with symbolic expressions and updates (conjoins) the path condition, while discarding unsat branches. Then, we define symbolic execution as a function $\mathrm{symex}:\mathcal P\times \widehat\Sigma \to 2^{\widehat\Sigma}$ such that
\[
\mathrm{symex}(P,\hat\sigma)\;=\;\{\hat\sigma' \in \widehat\Sigma \mid \langle P,\hat\sigma\rangle \Rightarrow^{*} \langle \_,\hat\sigma'\rangle\}.
\]

\begin{example}[Symbolic execution with a branch]
Let the program be $P \triangleq \texttt{if }(x > 0)\texttt{ then }y := x+1\texttt{ else }y := 0$. Assume we start from the symbolic state
\[
\hat\sigma_0=(\hat\nu_0,\hat h_0,\pi_0)
\quad\text{with}\quad
\hat\nu_0(x)=\alpha,\;\hat\nu_0(y)=\beta,\;\hat h_0=\emptyset,\;\pi_0=\top,
\]
where $\alpha,\beta\in X$ are fresh logical variables. Symbolic execution explores both branches, adding the corresponding guard to the path condition $\langle P,\hat\sigma_0\rangle \Rightarrow^{*} \langle \_,\hat\sigma_1\rangle$ and $\langle P,\hat\sigma_0\rangle \Rightarrow^{*} \langle \_,\hat\sigma_2\rangle$, where
\begin{align*}
&\hat\sigma_1 = (\hat\nu_1,\emptyset,\pi_1),
\quad
\pi_1 \equiv \alpha>0,
\quad
\hat\nu_1(y)=\alpha+1,\;\hat\nu_1(x)=\alpha,\\
&\hat\sigma_2 = (\hat\nu_2,\emptyset,\pi_2),
\quad
\pi_2 \equiv \alpha\le 0,
\quad
\hat\nu_2(y)=0,\;\hat\nu_2(x)=\alpha.
\end{align*}
Thus, $\mathrm{symex}(P,\hat\sigma_0)=\{\hat\sigma_1,\hat\sigma_2\}$.
\end{example}

\section{\projName}

\begin{figure}[t]
    \centering
        {\scriptsize
      \begin{tcolorbox}[mybox,title={Ghost Inversion Prompt Template}]
        \begin{minted}[escapeinside=||]{text}
Generate a function {SIGN} that inverts the original code fragment to recover the pre-state values of the
variables {VARS}. If exact inversion is not possible, implement a bounded numeric search or a clamped
approximation, and clearly comment on this in the output. The original function code: {CODE}
      \end{minted}
      \end{tcolorbox}
        }
\vspace{-2mm}
\caption{\label{fig:inverse_template}A simplified fragment of \projName's prompt template for generating ghost code that inverts a difficult fragment, where \texttt{SIGN} is the signature of the desired function, \texttt{VARS} are the variables, whose values the inversion needs to compute, and \texttt{CODE} is the code fragment.}
\vspace{-2mm}
\end{figure}
\projName{} adds a layer on top of a traditional SMT-backed symbolic execution engine. 
\rev{\projName{} first performs an exploratory KLEE run with solver timeouts reduced by 90\% to identify uncovered branches, which mark hard-to-reach paths. These candidate branches are then ranked by how high they occur in the control flow and by the LOC size of the region they guard, prioritizing branches whose resolution could unlock larger still-uncovered code regions. For each branch, \projName{} constructs the LLM input \{CODE\} from the branch condition, the enclosing function, preceding code that may influence it, and directly related callees, truncated to a fixed context budget.} Based on this fragment
and the LLM's choice, \projName{} generates one of three kinds of ghost code to improve constraint solving and path exploration: (1) fragment inversion, (2) surrogate models, and (3) semantic heap partitioning. 
\rev{The process is fully automated, not requiring manual hints or hand-crafted inversions.} 
The ghost code is then integrated into the symbolic executor and solver as explained below.

\subsection{Bidirectional Constraint Propagation with Code Inversion}
\label{sec:bidirectional}

\begin{figure}[t]
  \begin{subfigure}[b]{0.48\textwidth}
    \centering
{\scriptsize          
\begin{minted}[escapeinside=||]{C}
int hard_func(int x) { 
    return (x * x + 6) % 97; 
}
int main(void) {
    int a = |$\alpha$|;
    |\hly{int b = hard\_func(a);}|
    if (b == 42) {
        assert(a > 20);
    }
}
\end{minted}
}
  \end{subfigure}
  \hfill
  \begin{subfigure}[b]{0.48\textwidth}
    \centering
{\scriptsize          
\begin{minted}[escapeinside=||]{C}
void f_inv(int b, int *results) {
    int count = 0;
    int t = (b - 6) % 97;
    if (t < 0) t += 97;
    for (int a = 0; a < 97; a++) {
        if ((a * a) % 97 == t) {
            results[count++] = a;
            if (count == 2) break;
        }
    }
}
\end{minted}
}
  \end{subfigure}
  \vspace{-2mm}
  \caption{Running example to illustrate ghost inversion and bidirectional constraint propagation.\label{fig:inversion_running_example}} 
  \vspace{-2mm} 
\end{figure}

Given a program $P$ in which an LLM has identified a fragment $C$ as difficult for SMT reasoning, e.g. non-linear arithmetic, \projName uses an LLM to generate a ghost inversion procedure. Intuitively, we treat $C$ as a (partial) state transformer on a selected slice of the state --- i.e., the set of program variables and heap locations that $C$ may read or write --- and synthesize code that implements an inverse of that transformer. Executing the ghost inversion in reverse then propagates constraints from the post-state back to the pre-state, allowing the solver to reason about $C$ indirectly even when its forward semantics is not handled well by the underlying theory.

Let the sets of program variables and heap locations that a code fragment \(C\) may read/write be
\[
\mathsf{RdVar}(C),\mathsf{WrVar}(C)\subseteq \mathsf{Var},\qquad
\mathsf{RdLoc}(C),\mathsf{WrLoc}(C)\subseteq \mathrm{Addr}.
\]
Then, the read/write footprints of \(C\) are defined as
\[
\mathsf{Rd}(C)\;\triangleq\;\mathsf{RdVar}(C)\cup \mathsf{RdLoc}(C),
\qquad
\mathsf{Wr}(C)\;\triangleq\;\mathsf{WrVar}(C)\cup \mathsf{WrLoc}(C),
\]
and we denote \(\sigma\!\upharpoonright_{S}\) for projecting a concrete state \(\sigma\) to the values of a set \(S\) of variables/heap locations. We model \(C\) as a function from the values $C$ reads to the values it writes, $f_C \;:\; \Sigma_{\mathsf{Rd}(C)} \rightarrow \Sigma_{\mathsf{Wr}(C)}$:
\[
f_C(\rho)=\omega
\;\;\triangleq\;\;
\exists \sigma.\;
\Bigl(\rho=\sigma\!\upharpoonright_{\mathsf{Rd}(C)}
\;\wedge\;
\omega=\mathrm{exec}(C,\sigma)\!\upharpoonright_{\mathsf{Wr}(C)}\Bigr),
\]
\noindent Using the prompt shown in \Cref{fig:inverse_template}, the LLM is requested to generate a ghost function \mintinline{C}`f\_inv` that implements an inverse $f^{-1}_C \;:\; \Sigma_{\mathsf{Wr}(C)} \rightharpoonup \Sigma_{\mathsf{Rd}(C)}$ such that
\[
\forall \rho\in \mathrm{Dom}(f_C).\;\;
f^{-1}_C\!\bigl(f_C(\rho)\bigr)=\rho.
\]
Operationally, this means that for any concrete state \(\sigma\) where both executions terminate,
\[
\mathrm{exec}(\mathtt{f\_inv},\,\mathrm{exec}(C,\sigma))\!\upharpoonright_{\mathsf{Rd}(C)}
\;=\;
\sigma\!\upharpoonright_{\mathsf{Rd}(C)}.
\]

In practice, many fragments cannot be fully inverted, and an LLM is instead requested to return a set of possible values that are all valid inverses, i.e. $\forall \rho\in \mathrm{Dom}(f_C).\;\; \rho \in f^{-1}_C\!\bigl(f_C(\rho)\bigr)$, or when no inverse values exist, the generated function is tasked to find an input value that results in the closest output state to the expected state, as shown in the example in \Cref{fig:inversion}.

\begin{example}[Ghost inversion for a hard fragment]
  Consider the program fragment in \Cref{fig:inversion_running_example} (left). Let $C$ be the call/statement $b := \mathsf{hard\_func}(a)$. Its read/write sets are 
\[
\mathsf{RdVar}(C)=\{a\},\qquad \mathsf{WrVar}(C)=\{b\},\qquad \mathsf{RdLoc}(C)=\mathsf{WrLoc}(C)=\emptyset,
\]
hence $\mathsf{Rd}(C)=\{a\}$ and $\mathsf{Wr}(C)=\{b\}$. For this fragment, an LLM generates the function \mintinline{C}`f_inv` shown on the right of the same figure, where the output of the function is the list of values \mintinline{C}`results`.
\end{example}
  
\projName replaces $C$ with an assignment of $\mathsf{Wr}(C)$ to fresh symbolic variables, since it is not able to directly reason about the output of $C$ using an SMT solver. Assume \(\mathsf{Wr}(C)=\{\nu_1,\dots,\nu_k\}\), and \(\beta_1,\dots,\beta_k\in X\) be fresh logical symbols w.r.t. the current symbolic state. Let the havoc statement be
\[
\mathsf{havoc}(\mathsf{Wr}(C)) \;\triangleq\; (\nu_1 := \beta_1;\;\cdots;\; \nu_k := \beta_k).
\]
Then, \projName transforms the program by replacing $C$ with this havoc statement:
\[
P' \triangleq P[C \mapsto \mathsf{havoc}(\mathsf{Wr}(C))].
\]

Finally, \projName symbolically executes $P'$ from an initial symbolic state
$\hat\sigma_0=(\hat\nu_0,\hat h_0,\pi_0)$, which is $\hat h_0=\emptyset$, $\pi_0\equiv \top$, and $\hat\nu_0(a)=\alpha$ (fresh $\alpha\in X$) in the running example. Let the set of final symbolic states produced by symbolic execution of the transformed program be
\[
\mathrm{symex}(P',\hat\sigma_0)
\;=\;
\{\hat\sigma_1^1,\ldots,\hat\sigma_1^m\},
\]

\noindent where each \(\hat\sigma_1^i=(\hat\nu_1^i,\hat h_1^i,\pi_1^i)\) is a (feasible) terminal symbolic state. To make explicit what happens before and after the replaced fragment \(C\), we decompose each $\pi_1^i$ into corresponding prefix and suffix constraints: $\pi_1^i = \pi_\mathrm{pre}^i \wedge \pi_\mathrm{suf}^i$, where $\pi_\mathrm{pre}^i$ depends only on the initial input states, while $\pi_\mathrm{suf}^i$ also depends on the additional symbolic variables introduced by the havoc statement\footnote{The path condition has to be decomposed into a sequence of sub-constraints if $C$ is executed inside a loop.}. In the running example, the two final path conditions decompose as:
\begin{align*}
\pi_f^{(1)} \equiv (\,\beta = 42\,)\wedge(\,\alpha>20\,)
\;&=\;
\underbrace{\top}_{\pi_{\mathrm{pre}}^{(1)}}\wedge
\underbrace{((\beta=42)\wedge(\alpha>20))}_{\pi_{\mathrm{suf}}^{(1)}},\\
\pi_f^{(2)} \equiv (\,\beta \neq 42\,)
\;&=\;
\underbrace{\top}_{\pi_{\mathrm{pre}}^{(2)}}\wedge
\underbrace{(\beta\neq 42)}_{\pi_{\mathrm{suf}}^{(2)}}.
\end{align*}

For each pair of $\pi_\mathrm{pre}^i, \pi_\mathrm{suf}^i$, after computing the above path conditions, \projName executes our bidirectional constraint propagation algorithm (\Cref{alg:bidirectional_reconcile}) that reconciles prefix and suffix constraints around a transformed fragment \(f\) by iteratively ``meeting in the middle'' using the generated inverse \(f^{-1}\). It first solves the current suffix constraint \(\pi_{\mathrm{suf}}\) to obtain an initial assignment \(A_{\mathrm{suf}}\); from this it extracts proposed values for the variables written by \(f\) and runs \(f^{-1}\) to propagate them backward to a proposed pre-state \(\sigma_{\mathrm{pre}}\) over the variables read by \(f\). It then finds an assignment \(A_{\mathrm{pre}}\) satisfying \(\pi_{\mathrm{pre}}\) that is as close as possible (under a domain-specific distance discussed below) to \(\sigma_{\mathrm{pre}}\), propagates \(A_{\mathrm{pre}}\) forward through \(f\) to get an output state \(\sigma_{\mathrm{out}}\), and symmetrically optimizes within \(\pi_{\mathrm{suf}}\) to update \(A_{\mathrm{suf}}\) so that it is close to \(\sigma_{\mathrm{out}}\). If it satisfies \(\pi_{\mathrm{pre}}\wedge\pi_{\mathrm{suf}}\), it returns it; otherwise it repeats the loop until it reaches a fixpoint or a solver call on \(\pi_{\mathrm{pre}}\) or \(\pi_{\mathrm{suf}}\) returns \textsc{Unsat}. 
\rev{Under an approximate inverse, \textsc{Unsat} is operational: \projName{} found no satisfying assignment within the current approximation and budget; the path itself may or may not be feasible.}

The algorithm drives the prefix--suffix reconciliation toward a fixpoint by treating the values propagated across the cut as soft targets and enforcing consistency via optimization. Concretely, after executing \(f^{-1}\) (resp., \(f\)) we obtain a suggested valuation \(\sigma_{\mathrm{pre}}\) for the pre-variables (resp., \(\sigma_{\mathrm{out}}\) for the post-variables). We then choose, among all assignments satisfying the hard constraints \(\pi_{\mathrm{pre}}\) (resp., \(\pi_{\mathrm{suf}}\)), one that stays as close as possible to the suggested valuation.

When the interface between the two sides is a single numeric variable \(x\), we use Z3’s optimization engine~\cite{bjorner2015nuz} and minimize an \(\ell_1\) distance objective:
\[
A_{\mathrm{pre}} \;=\; \arg\min_{A \models \pi_{\mathrm{pre}}}\; |A(x)-\sigma_{\mathrm{pre}}(x)|,
\qquad
A_{\mathrm{suf}} \;=\; \arg\min_{A \models \pi_{\mathrm{suf}}}\; |A(x)-\sigma_{\mathrm{out}}(x)|.
\]
This makes each update the smallest possible correction needed to restore feasibility, which empirically stabilizes the iteration. When the interface consists of multiple variables \(V=\{x_1,\ldots,x_n\}\), optimizing a numeric distance can scale poorly. Instead, we bias updates toward changing as few variables as possible. We encode, for each \(x_i \in V\), a soft equality \(x_i=\hat x_i\) where \(\hat x_i\) is the suggested value from \(\sigma_{\mathrm{pre}}\) (or \(\sigma_{\mathrm{out}}\)), and solve a \emph{partial MaxSAT} instance with \(\pi_{\mathrm{pre}}\) (or \(\pi_{\mathrm{suf}}\)) as hard constraints:
\[
\max\;\sum_{i=1}^n \mathbf{1}\bigl(A(x_i)=\hat x_i\bigr)
\quad\text{s.t.}\quad A \models \pi_{\mathrm{pre}} \;\;(\text{resp. } \pi_{\mathrm{suf}}).
\]

\begin{wrapfigure}{r}{0.5\textwidth}
\resizebox{0.9\textwidth}{!}{%
  \begin{minipage}{\textwidth}
  \begin{algorithm}[H]
\KwIn{Prefix constraint $\pi_{\mathrm{pre}}$, suffix constraint $\pi_{\mathrm{suf}}$,\\difficult fragment $f$, synthesized inverse $f^{-1}$}
\KwOut{A model/assignment $A$ or \textsc{Unsat}}

\BlankLine
$\pi \gets \pi_{\mathrm{pre}} \wedge \pi_{\mathrm{suf}}$\;
$\mathsf{target}_{\mathrm{suf}} \gets \mathrm{vars}(\pi_{\mathrm{suf}})$\;
$\mathsf{target}_{\mathrm{pre}} \gets \mathrm{vars}(\pi_{\mathrm{pre}})$\;
$A_{\mathrm{suf}} \gets \mathrm{SMT}\bigl(\pi_{\mathrm{suf}}\bigr)$\;
\If{$A_{\mathrm{suf}}=\bot$}{\Return{\textsc{Unsat}}}

\BlankLine
\While{\textbf{true}}{

  $\sigma_{\mathrm{pre}} \gets \mathrm{exec}\bigl(f^{-1},\, A_{\mathrm{suf}}\!\upharpoonright_{\mathsf{Wr}(f)}\bigr)$\;

  $A_{\mathrm{pre}} \gets \mathrm{OPT}\Bigl(
      \mathrm{dist}\bigl(\mathsf{target}_{\mathrm{pre}},\, \sigma_{\mathrm{pre}}\bigr),
      \ \pi_{\mathrm{pre}}
    \Bigr)$\;
  \If{$A_{\mathrm{pre}}=\bot$}{\Return{\textsc{Unsat}}}

  $\sigma_{\mathrm{out}} \gets \mathrm{exec}\bigl(f,\, A_{\mathrm{pre}}\!\upharpoonright_{\mathsf{Rd}(f)}\bigr)$\;

  $A_{\mathrm{suf}} \gets \mathrm{OPT}\Bigl(
      \mathrm{dist}\bigl(\mathsf{target}_{\mathrm{suf}},\, \sigma_{\mathrm{out}}\bigr),
      \ \pi_{\mathrm{suf}}
    \Bigr)$\;
  \If{$A_{\mathrm{suf}}=\bot$}{\Return{\textsc{Unsat}}}

  \If{$A_{\mathrm{suf}} \models \pi_{\mathrm{pre}} \wedge \pi_{\mathrm{suf}}$}{
    \Return{$A_{\mathrm{suf}}$}
  }
}
  \end{algorithm}
  \end{minipage}
}
\vspace{-3mm}
\caption{Bidirectional constraint propagation.\label{alg:bidirectional_reconcile}}
\vspace{-3mm}
\end{wrapfigure}

In our running example, \projName uses the ghost inverse \mintinline{C}{f_inv} to enumerate candidate pre-values consistent with the observed havoc value $\{\,a \in \mathrm{Val} \mid a \in f^{-1}_C(42)\,\}=\{6,91\}.$ Finally, candidates are filtered by the remaining suffix (post-fragment) constraints over pre-variables (here, $\alpha>20$), leaving $a=91$ as the only value that satisfies the explored path. \Cref{fig:complex_math_example} shows a more complex example.

\begin{figure}[t]
    \centering
        {\scriptsize
      \begin{tcolorbox}[mybox,title={Ghost Model Prompt Template}]
        \begin{minted}[escapeinside=||]{text}
Rewrite the original fragment so that (1) functionality (outputs, branches) is preserved, (2) the rewritten
code translates to a solver-friendly SMT theory such as bitvectors or linear arithmetic. For any external
library or crypto call, provide a stub implementation. The original code fragment is {CODE}.
      \end{minted}
      \end{tcolorbox}
        }
\vspace{-3mm}
\caption{\label{fig:model_template}A fragment of \projName's prompt template for generating ghost model for a difficult fragment in \texttt{CODE}.}
\vspace{-4mm}
\end{figure}

\subsection{Surrogate Models}
\label{sec:surrogate-models}

Some difficult fragments are poor candidates for inversion: they may discard information (many-to-one), or depend on external state. In such cases, \projName asks an LLM to synthesize a \emph{surrogate model} --- a solver-friendly replacement that preserves the fragment's relevant behavior while avoiding solver-hostile theories.
Let \(C\) be the difficult fragment, with read/write footprints
\(\mathsf{Rd}(C)\) and \(\mathsf{Wr}(C)\), modeled as a state transformer
\(f_C:\Sigma_{\mathsf{Rd}(C)} \to \Sigma_{\mathsf{Wr}(C)}\).
Using the prompt template in \Cref{fig:model_template}, \projName{} queries an
LLM to generate a ghost fragment \(C_{\mathsf{sur}}\) implementing a surrogate
transformer \(g_C:\Sigma_{\mathsf{Rd}(C)} \to \Sigma_{\mathsf{Wr}(C)}\), which
approximates \(f_C\) while yielding solver-friendly constraints.
The program is rewritten as:
\[
P' \triangleq P[C \mapsto C_{\mathsf{sur}}],
\qquad
\mathrm{symex}(P',\hat\sigma_0)=\{\hat\sigma_1^1,\ldots,\hat\sigma_1^m\}.
\]
Each terminal symbolic state
\(\hat\sigma_1^i=(\hat\nu_1^i,\hat h_1^i,\pi_1^i)\) induces a simplified path
condition \(\pi_1^i\) corresponding to the original program \(P\).
Because \(g_C\) may be approximate, \projName treats it as a heuristic aid rather than a trusted semantic replacement. Any concrete input produced by solving constraints from \(P'\) is validated by re-executing the original program \(P\) to ensure the intended path is feasible in \(P\). \rev{This guarantees soundness of reported coverage but not completeness: approximate inversions, surrogate imprecision or early non-convergence may cause \projName{} to miss feasible paths.}

\begin{figure}[t]
    \centering
        {\scriptsize
      \begin{tcolorbox}[mybox,title={Ghost Topology Builder Prompt Template}]
        \begin{minted}[escapeinside=||]{text}
Identify a small number of distinct heap or pointer configurations that drive different control-flow outcomes.
Generate a function {SIGN} that builds representative memory shapes and pointer configurations. Leave concrete
values in the data structures symbolic. Code context: {CODE}
      \end{minted}
      \end{tcolorbox}
        }
\vspace{-3mm}
\caption{\label{fig:topology_template}A fragment of \projName's prompt template for generating ghost code that constrains heap, where \texttt{SIGN} is the expected function signature, \texttt{CODE} is a heap manipulating function, and data structure definitions.}
\vspace{-3mm}
\end{figure}

\subsection{Semantic Heap Partitioning}

Symbolic execution often fails on pointer-rich code because many possible aliasing patterns and heap shapes cause path explosion. \projName mitigates this by generating ghost code that partitions the unbounded heap space into a small set of \emph{semantic topologies} --- partially concrete, partially symbolic heap configurations corresponding to meaningful shapes in the program's domain. Such ghost code is generated automatically for all functions that operate on unbounded symbolic pointers.

\begin{example}[List segment of length $\ge 3$.]\label{example:topology}
Let $\mathsf{head},\mathsf{n}_1,\mathsf{n}_2 \in \mathsf{Var}$ be program variables of pointer type, and $\mathsf{next}$ and $\mathsf{val}$ be data fields. Let $X$ contain address-typed symbols $\alpha_0,\alpha_1,\alpha_2,\alpha_3$ and value-typed symbols $\beta_0,\beta_1,\beta_2$. We define a \emph{semantic topology} that represents the shape
\[
\alpha_0 \xrightarrow{\mathsf{next}} \alpha_1 \xrightarrow{\mathsf{next}} \alpha_2 \xrightarrow{\mathsf{next}} \alpha_3
\]
where $\alpha_0,\alpha_1,\alpha_2$ are the first three concrete links, node values $\beta_0,\beta_1,\beta_2$ are symbolic, and the tail pointer $\alpha_3$ remains symbolic (so the list has length $\ge 3$, with an unconstrained remainder). This semantic topology is represented via the symbolic state $\hat\sigma=(\hat\nu,\hat h,\pi)$, where the first three nodes are \emph{concretely} linked, while the tail pointer is symbolic:
\begin{align*}
&\hat\nu(\mathsf{head}) = \alpha_0,\qquad
\hat\nu(\mathsf{n}_1) = \alpha_1,\qquad
\hat\nu(\mathsf{n}_2) = \alpha_2, \qquad \hat h(\alpha_0) = \langle \mathsf{val} \mapsto \beta_0,\;\mathsf{next} \mapsto \alpha_1\rangle,\\
&\hat h(\alpha_1) = \langle \mathsf{val} \mapsto \beta_1,\;\mathsf{next} \mapsto \alpha_2\rangle,\quad
\hat h(\alpha_2) = \langle \mathsf{val} \mapsto \beta_2,\;\mathsf{next} \mapsto \alpha_3\rangle.
\end{align*}
The path condition $\pi$ enforces that the three nodes are distinct, allocated, and that $\alpha_3$ denotes an arbitrary (possibly null) tail:
\[
\pi \triangleq (\alpha_0 \neq \mathsf{null}) \wedge (\alpha_1 \neq \mathsf{null}) \wedge (\alpha_2 \neq \mathsf{null})
\wedge (\alpha_0 \neq \alpha_1) \wedge (\alpha_1 \neq \alpha_2) \wedge (\alpha_0 \neq \alpha_2).
\]
\end{example}

\projName adds heap-topology constraints by executing generated ghost code. The key idea is to (i) introduce
a fresh selector variable $\xi$ and (ii) conditionally constrain the symbolic
heap to satisfy one of a small set of semantic topologies identified based on the definition of the data structure, and the code context. Each topology yields a solver-friendly, bounded set of constraints.

Let the fragment \(C\) be a call site \(\mathsf{call}\;p(r)\) where the argument
\(r\) is pointer-typed and fully symbolic. Using the prompt template in
\Cref{fig:topology_template}, \projName{} asks the LLM to synthesize a ghost
procedure \(\mathsf{constrain\_heap}_p(r,\xi)\), where \(\xi \in X\) is a fresh
symbolic variable selecting among candidate heap topologies. The program is then
rewritten as \(P' \triangleq P[C \mapsto \mathsf{constrain\_heap}_p(r,\xi);\,C]\),
inserting the ghost call immediately before \(C\).
Symbolic execution then proceeds on $P'$, where the added ghost code restricts
the path condition to heaps consistent with the chosen topology; other
topologies are explored by varying $\xi$.

\begin{wrapfigure}{r}{0.62\textwidth}
\vspace{-4mm}
{\scriptsize
\begin{minted}[escapeinside=||]{C}
void constrain_heap(Node* a0, Node* a1, Node* a2, Node* a3, int xi) {
  if (xi == 0) { /* list segment of length >= 3 */
    assume(a0 != NULL);
    assume(a1 != NULL && a2 != NULL);
    assume(a0 != a1 && a1 != a2 && a0 != a2);
    assume(a0->val == |$\beta_0$|);  assume(a0->next == a1);
    assume(a1->val == |$\beta_1$|);  assume(a1->next == a2);
    assume(a2->val == |$\beta_2$|);  assume(a2->next == a3);
  }
  /* else if (xi == 1) { ... other topologies ... } */
}
\end{minted}
}
\vspace{-4mm}
\end{wrapfigure}
\begin{example}[Ghost code for the list-topology of length $\ge 3$]
Suppose we call a procedure \texttt{foo} with a fully symbolic list head $\mathsf{call}\;\texttt{foo}(\mathsf{head}).$ We insert a ghost call right before it:
\[
\mathsf{constrain\_heap}_{\texttt{foo}}(\mathsf{a0},\mathsf{a1},\mathsf{a2},\mathsf{a3},\xi);\;
\mathsf{call}\;\texttt{foo}(\mathsf{head}).
\]
Below is one branch of $\mathsf{constrain\_heap}_{\texttt{foo}}$ implementing
the topology from Example~\ref{example:topology}. The ghost code links three concrete nodes, while leaving the list tail and the values unconstrained.
\end{example}

\begin{table}[t!]
\centering
\caption{Results on the LogicBombs benchmark.
Each program contains exactly two critical paths (bomb-triggering and benign).
Left: KLEE baselines.
Right: LLM-based tools.
Tokens are reported in millions.
The total number of bombs is 53 and the total number of critical paths (CPs) is
$53 \times 2 = 106$.}
\label{tab:logic-bombs-summary}
\footnotesize
\vspace{-2mm}
\begin{tabular}{@{}p{0.28\columnwidth}
@{\hspace{0.00\columnwidth}}
p{0.70\columnwidth}@{}}

\raggedright
\centering\textbf{KLEE Baselines}\par\vspace{2pt}
\centering
\begin{tabular}{|l|cc|}
\toprule
\multicolumn{3}{|c|}{\rule{0pt}{3.2ex}} \\  
\textbf{Solver}
& \makecell{\textbf{Bombs}\\\textbf{Trig.}}
& \makecell{\textbf{CPs}\\\textbf{Cov.}} \\
\midrule
STP        & 20 & 59 \\
Z3         & 17 & 56 \\
Float-Z3   & 21 & 60 \\
\bottomrule
\end{tabular}

&
\centering\textbf{LLM-Based Tools}\par\vspace{2pt}
\centering
\begin{tabular}{|l|ccc|ccc|}
\toprule
\multicolumn{1}{|c|}{\textbf{Backend}}
& \multicolumn{3}{c|}{\textbf{\projName{}}}
& \multicolumn{3}{c|}{\textbf{ConcoLLMic}} \\
\cmidrule(lr){2-4}\cmidrule(lr){5-7}
& \makecell{\textbf{Bombs}\\\textbf{Trig.}}
& \makecell{\textbf{CPs}\\\textbf{Cov.}}
& \makecell{\textbf{Tokens}}
& \makecell{\textbf{Bombs}\\\textbf{Trig.}}
& \makecell{\textbf{CPs}\\\textbf{Cov.}}
& \makecell{\textbf{Tokens}} \\
\midrule
Claude 3.7
& \textbf{50} & \textbf{103} & 0.30
& 35 & 83 & 4.03 \\

GPT-5
& 49 & 102 & 0.32
& 33 & 81 & 2.14 \\

DeepSeek-Chat
& 45 & 98 & 0.33
& 34 & 82 & 3.04 \\
\bottomrule
\end{tabular}

\end{tabular}
\vspace{-4mm}
\end{table}

\rev{Semantic heap partitioning is a heuristic: it biases exploration toward a finite set of meaningful topologies and may miss behaviors whose required topology is not generated or explored within the budget. This can yield false negatives but not false positives.}

\section{Evaluation}
\label{sec:evaluation}

Our experiments address the following research questions:
\begin{description}
\item[RQ1:] Do LLM-generated ghost code techniques in \projName unlock solver-hostile paths and improve coverage compared to existing symbolic and LLM-based approaches?
\item[RQ2:] Does \projName scale to real-world numerical and structured-input programs?
\item[RQ3:] How does LLM token cost of \projName compare to prior LLM-based approaches?
\item[RQ4:] \rev{How quickly does \projName{} accumulate coverage relative to baselines?}
\item[RQ5:] \rev{Does \projName{} explore hard-to-reach paths that other techniques miss?}
\end{description}

\paragraph*{Experimental Setup} \projName is implemented on top of KLEE~\cite{klee} with Z3~\cite{de2008z3}, and KLEE-Float was used on instances involving floating point numbers. All experiments were conducted on a single Linux machine running Ubuntu 24.04.3 LTS with an Intel® Core™ Ultra 7 258V CPU and 32~GiB of RAM. All symbolic execution runs were conducted inside Docker containers to provide isolation and reproducibility. Each tool configuration was allocated a fixed wall-clock budget of 36 hours, which was applied uniformly across all tools, baselines, and ablations, following standard practice in symbolic execution and software testing evaluations.

\paragraph*{LLM Backends.} All LLM inference was performed via external APIs; we used three state-of-the-art LLMs from different families: (1) \textbf{GPT-5} (version \texttt{gpt5-2025-08-07})~\cite{openai_gpt5}, (2) \textbf{Claude-3.7} (version \texttt{claude-3-7-sonnet-20250219})~\cite{anthropicclaude}, and (3) \textbf{DeepSeek-Chat} (version 3.2, release 2025-12-01)~\cite{deepseekv3}. We used the sampling temperature of 0.5, a standard mid-range setting, and to enhance reproducibility, we cached all LLM responses so that they can be replayed deterministically~\cite{dai2025statisticalindependenceawarecaching}.


\paragraph*{Benchmarks} We used three benchmarks to stress different symbolic execution challenges:
\begin{itemize}
\item \textbf{LogicBombs} benchmark suite introduced by Xu et al.~\cite{xu2018benchmarking} consists of small, synthetic C programs embedding guarded ``logic bombs'', each targeting a specific symbolic execution challenge (e.g., arithmetic reasoning, floating-point conditions, symbolic memory, and path explosion). In total, it comprises 53 programs with 1{,}121 lines of executable code; each of them isolates a single challenge, enabling controlled analysis of which reasoning techniques are effective and how individual components of \textsc{\projName{}} contribute to the overall result.
\item \textbf{FDLibM} (Freely Distributable LibM)~\cite{FDLibM} is a popular math library involving complex arithmetic and floating-point guards. Our evaluation covers 78 entry points, exercising different parts of the library, totaling 5{,}013 lines of executable code. FDLibM contains deeply nested control flow, non-linear arithmetic, and floating-point branching that frequently leads to concretization or solver failure in conventional symbolic execution. This benchmark therefore allows us to evaluate \projName's ability to reason about inversion patterns, mathematical structure, and floating-point edge cases in real-world numerical code.
 \item \textbf{Structured-Input Programs}: to evaluate \projName capabilities on larger, real-world codebases, we consider programs that process structured input, since such programs rely on complex dynamic data structures. We chose \rev{four} widely-used programs: (1) \texttt{libexpat} (XML parsing, 11{,}828  LOC), (2) \texttt{jq} (JSON processing, 3{,}399  LOC), (3) GNU \texttt{bc} (POSIX expression language, 8{,}182  LOC), \begin{revblock} and (4)  \texttt{libyaml} (YAML parsing, 2{,}760  LOC).   \end{revblock}
\end{itemize}

\begin{figure}[t]
  \centering
 \includegraphics[
  width=0.9\textwidth,
  trim=0 8 0 0,
  clip
 ]{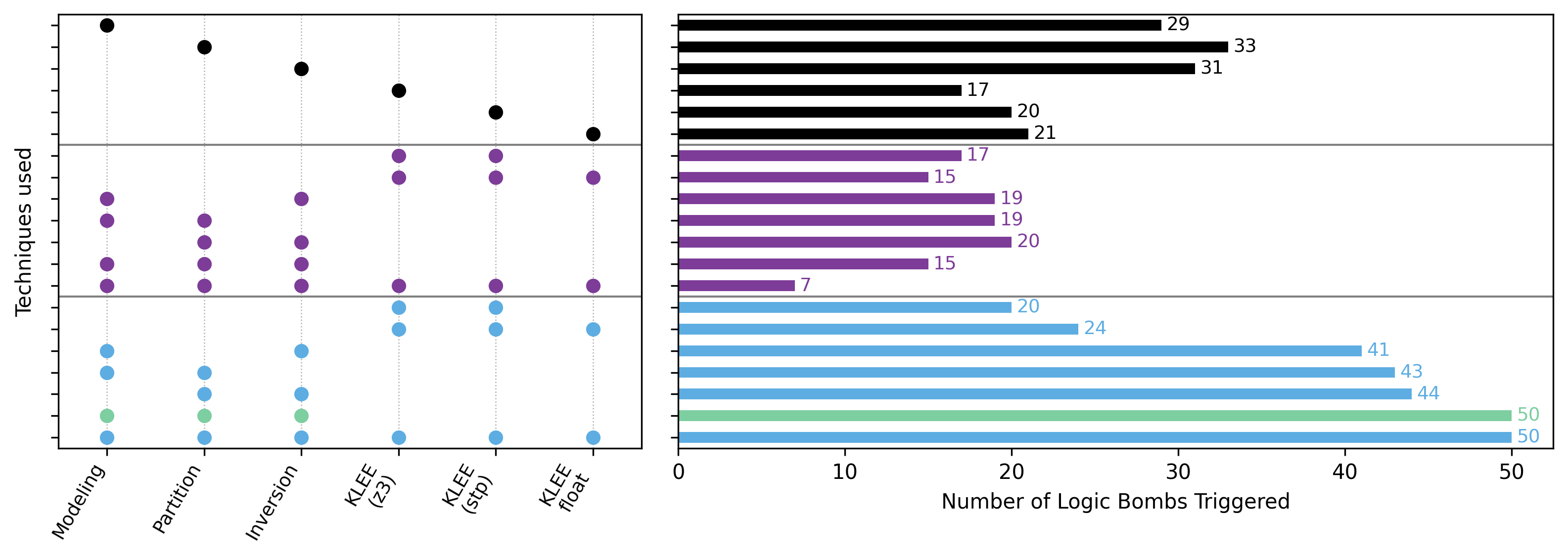}
 \vspace{-3mm}
  \caption{Coverage across different combinations of tool components. 
  Dots indicate enabled techniques, while bars report the number of logic bombs triggered.
  Black rows show individual techniques, purple rows show intersections (AND),
  blue rows show unions (OR), and the green row highlights the union of the three
  LLM-based techniques, which is equivalent to \projName.}
  \label{fig:technique-queries}
 \vspace{-3mm}
\end{figure}

\paragraph*{Baselines} We compared \projName against the following baselines:
\begin{itemize}
\item Traditional symbolic execution: (1) \textbf{KLEE 3.0 (STP)}, using the official \texttt{klee:3.0} Docker image with the STP solver, (2) \textbf{KLEE 3.0 (Z3)}, using the same image with Z3 as the backend solver, and (3) \textbf{KLEE-Float (Z3)}, using a fork of KLEE \cite{kleeFloat} designed for precise floating-point reasoning. All tools used the default settings provided by the corresponding distributions, using exactly the same options when KLEE serves as the \projName backend. \rev{Thus, unless explicitly stated otherwise, \projName{} uses the same KLEE/Z3 options as the corresponding KLEE baseline.}
\item \textbf{ConcoLLMic}~\cite{concollmic} as a representative LLM-based concolic execution baseline. It first performs code instrumentation, followed by LLM-guided test generation and path exploration. \rev{ConcoLLMic's instrumentation phase can consume much or all of the time budget before any test generation begins. To objectively assess its exploration capability, we exclude ConcoLLMic's instrumentation time from the 36-hour budget---a choice that favors ConcoLLMic over \projName{}.} We evaluated it using the same three LLMs: GPT-5, Claude-3.7, and DeepSeek-Chat.
\item \textbf{Cottontail}~\cite{cottontail} is an LLM-based concolic executor designed for grammar-constrained input generation. Thus, we compared its performance with \projName on the structured-input programs (\texttt{libexpat}, \texttt{jq}, \texttt{bc}, \texttt{libyaml}) using the same LLMs. We used JSON and XML grammars provided by the original implementation; and added grammar support for \texttt{bc} and YAML.
\end{itemize}

\subsection{RQ1: \projName's Ability To Unlock Solver-Hostile Paths}

\begin{table}[t!]
\centering
\caption{Results on the FDLibM benchmark.
Left: KLEE baselines.
Right: LLM-based tools.
\textbf{Avg} denotes mean per-function line coverage;
\textbf{Tot} denotes overall line coverage across all functions.
Tokens are in millions.}
\label{tab:FDLibM-summary}
\footnotesize
\vspace{-2mm}
\begin{tabular}{@{}p{0.28\columnwidth}
@{\hspace{0.00\columnwidth}}
p{0.70\columnwidth}@{}}

\raggedright
\centering\textbf{KLEE Baselines}\par\vspace{2pt}
\centering
\begin{tabular}{|l|cc|}
\toprule
\multicolumn{3}{|c|}{\rule{0pt}{3.2ex}} \\   
\textbf{Solver}
& \textbf{Avg}
& \textbf{Tot} \\
\midrule
STP        & 62.68 & 70.21 \\
Z3         & 63.09 & 68.98 \\
Float-Z3   & 56.67 & 57.32 \\
\bottomrule
\end{tabular}

&
\centering\textbf{LLM-Based Tools}\par\vspace{2pt}
\centering
\begin{tabular}{|l|ccc|ccc|}
\toprule
\multicolumn{1}{|c|}{\textbf{Backend}}
& \multicolumn{3}{c|}{\textbf{\projName{}}}
& \multicolumn{3}{c|}{\textbf{ConcoLLMic}} \\
\cmidrule(lr){2-4}\cmidrule(lr){5-7}
& \textbf{Avg}
& \textbf{Tot}
& \makecell{\textbf{Tokens}}
& \textbf{Avg}
& \textbf{Tot}
& \makecell{\textbf{Tokens}} \\
\midrule
Claude 3.7
& 91.14 & 83.62 & 0.49
& 40.02 & 31.14 & 16.45 \\

GPT-5
& \textbf{91.67} & \textbf{90.23 } & 0.54
& 38.65 & 30.62 & 10.53 \\

DeepSeek-Chat
& 91.12 & 83.31 & 0.50
& 37.17 & 28.98 & 8.78 \\
\bottomrule
\end{tabular}
\end{tabular}
\vspace{-4mm}
\end{table}

We compared \projName{} against KLEE baselines and ConcoLLMic on LogicBombs, which is designed to exercise diverse symbolic execution challenges. Each of its 53 programs has two critical paths---a logic-bomb path and a benign one---and we report bombs triggered and critical paths covered.

Table~\ref{tab:logic-bombs-summary} reports the results. KLEE triggers only 32.1--39.6\% of bombs and covers at most 56.6\% of critical paths, since solver-hostile constructs (non-linear arithmetic, floating-point conditions, complex data dependencies) leave many bomb paths unexplored. ConcoLLMic raises these to 62.3--66.0\% and 76.4--78.3\% by reasoning over execution semantics rather than strict constraints. \projName{} performs best, triggering up to 94.3\% of bombs and covering 97.2\% of critical paths---a 138.1\%/71.7\% gain over the best KLEE variant and 42.9\%/24.1\% over the best ConcoLLMic run.

\begin{figure}[t]
    \centering
    \includegraphics[width=0.85\linewidth]{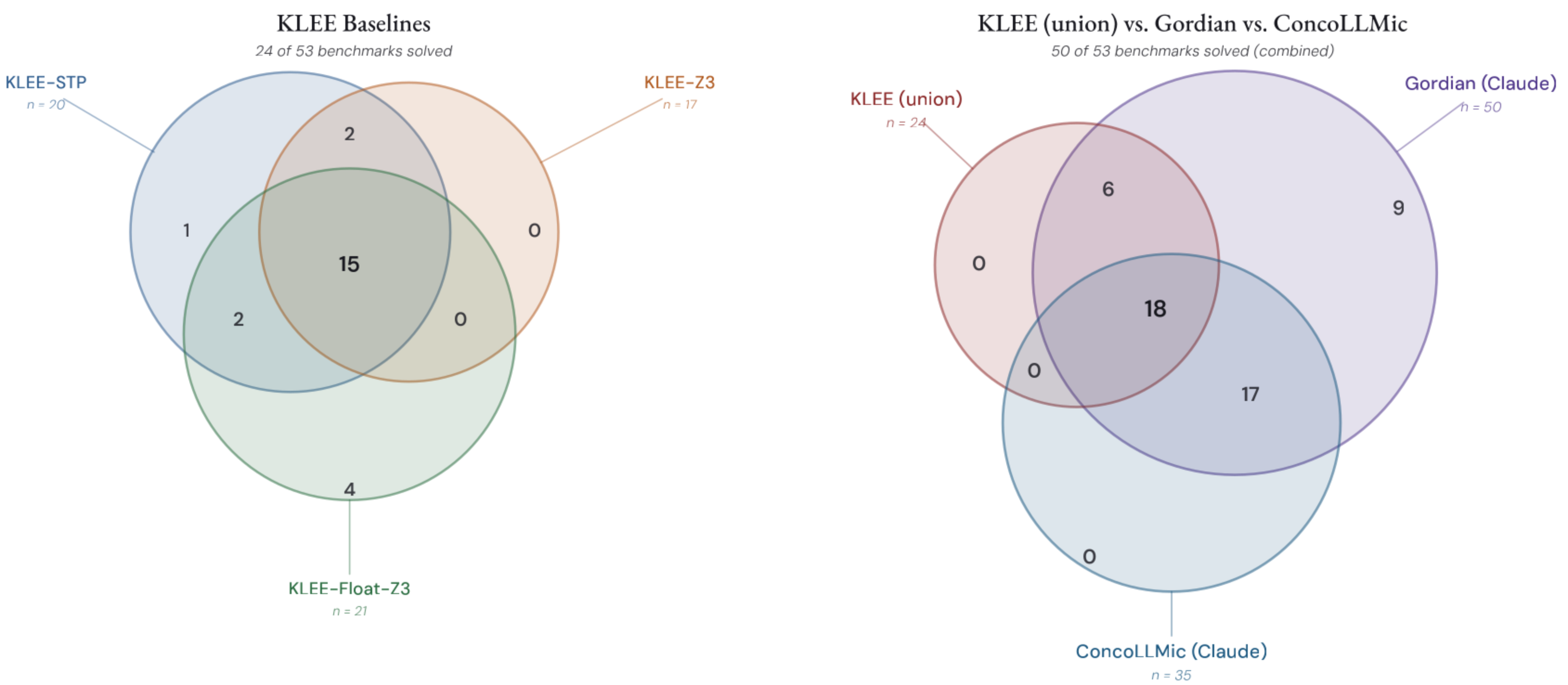}
    \vspace{-2mm}
    \caption{\rev{Overlap in bombs KLEE, \projName and ConcoLLMic trigger.\label{fig:bomb-venn}}}
    \vspace{-3mm}
\end{figure}

To trace this gain, Figure~\ref{fig:technique-queries} breaks coverage down by component, using Claude for the three ghost code techniques (\emph{Modeling}, \emph{Inversion}, \emph{Partition}). Each triggers a substantial but distinct subset of bombs, and their union (green row, equivalent to \projName{}) covers nearly all of them; their intersection is small, so few bombs need all three. The KLEE variants, in contrast, largely overlap. Adding KLEE to the ghost code union yields no further coverage---\projName{} is strictly additive over KLEE.

\rev{Figure~\ref{fig:bomb-venn} shows the overlap in bombs each approach triggers. The three KLEE solver variants agree closely, sharing 15 bombs with only marginal solver-specific differences at the margins. Both LLM-based approaches substantially extend KLEE's reach: all share common 18 bombs, beyond which \projName{} and ConcoLLMic jointly solve 17 more, and \projName{} exclusively solves 9. 
  The 50 bombs solved by \projName{} form a strict superset of those solved by KLEE and ConcoLLMic combined, showing both that LLM-augmented techniques extend traditional symbolic execution and that \projName{} covers all bombs reached by the other evaluated approaches on this benchmark.}

\begin{tcolorbox}[
    colback=gray!5,
    colframe=black!20,
    arc=2pt,
    boxrule=0.3mm,
    sharp corners,
    left=2pt,
    right=2pt,
    top=2pt,
    bottom=2pt,
]
\textbf{RQ1:} \projName{}'s LLM-generated ghost code unlocks solver-hostile paths that are unreachable by prior approaches, triggering up to \textbf{94.3\%} of logic bombs and covering \textbf{97.2\%} of critical paths, strictly subsuming baseline-triggered bombs. This corresponds to a \textbf{42.9\%} improvement over the strongest LLM-based baseline and a \textbf{138.1\%} improvement over the best KLEE configuration.
\end{tcolorbox}




\begin{table}[ht!]
\centering
\caption{Line coverage (\%) results on the structured-input programs.
Left: KLEE baselines.
Right: LLM-based tools.
Avg denotes mean per-file coverage; Tot denotes overall coverage.}
\label{tab:coverage-summary}
\footnotesize
\vspace{-4mm}
\hspace*{-8mm}\begin{tabular}{@{}p{0.36\columnwidth}
@{\hspace{0.00\columnwidth}}
p{0.56\columnwidth}@{}}

\raggedright
\centering\textbf{KLEE Baselines}\par\vspace{2pt}
\centering
\begin{tabular}{|l|lcc|}
\toprule
\multicolumn{4}{|c|}{\rule{0pt}{3.2ex}} \\  
\textbf{Bench} & \textbf{Solver} & \textbf{Avg} & \textbf{Tot} \\
\midrule
\multirow{3}{*}{libexpat}
 & Z3        & 20.21 & 14.35 \\
 & STP       & 18.10 & 12.62 \\
 & Float-Z3  & 18.40 & 12.89 \\
\midrule
\multirow{3}{*}{jq}
 & Z3        & 34.29 & 28.95 \\
 & STP       & 35.09 & 29.80 \\
 & Float-Z3  & 30.81 & 25.34 \\
\midrule
\multirow{3}{*}{bc}
 & Z3        & 50.69 & 40.01 \\
 & STP       & 54.59 & 44.70 \\
 & Float-Z3  & 18.01 & 9.84 \\
  
 \midrule

\multirow{3}{*}{\rev{libyaml}}
 & \rev{Z3}        & \rev{21.03} & \rev{21.17} \\
 & \rev{STP}       & \rev{23.67} & \rev{23.74} \\
 & \rev{Float-Z3}  & \rev{15.18} & \rev{13.85} \\
\midrule
\midrule
\multirow{3}{*}{\rev{Overall}}
& \rev{Z3}        & \rev{32.99} & \rev{24.97} \\
& \rev{STP}       & \rev{34.31} & \rev{26.05} \\
& \rev{Float-Z3}  & \rev{20.38} & \rev{13.37} \\

\bottomrule
\end{tabular}

&
\centering\textbf{LLM-Based Tools}\par\vspace{2pt}
\centering
\begin{tabular}{|l|cc|cc|cc|}
\toprule
\multicolumn{1}{|c|}{\textbf{Backend}}
 & \multicolumn{2}{c|}{\textbf{\projName{}}}
 & \multicolumn{2}{c|}{\textbf{ConcoLLMic}}
 & \multicolumn{2}{c|}{\textbf{Cottontail}} \\
\cmidrule(lr){2-3}\cmidrule(lr){4-5}\cmidrule(lr){6-7}
 & \textbf{Avg} & \textbf{Tot}
 & \textbf{Avg} & \textbf{Tot}
 & \textbf{Avg} & \textbf{Tot} \\
\midrule
Claude 3.7    & 46.29 & 38.63 & 11.27 & 4.18 & 11.42 & 11.38 \\
GPT-5         & 59.72 & \textbf{47.75} & 13.04 & 5.11 & 21.86 & 18.30 \\
DeepSeek-Chat & 49.68 & 43.03 & 8.80 & 2.37 & 12.32 & 13.13 \\
\midrule
Claude 3.7    & 46.19 & \textbf{45.37} & 9.68 & 8.37 & 30.56 & 29.37 \\
GPT-5         & 48.69 & 40.98 & 9.75 & 8.56 & 35.34 & 31.74 \\
DeepSeek-Chat & 56.49 & 44.86 & 12.69 & 10.00 & 31.54 & 30.38 \\
\midrule
Claude 3.7    & 70.74 & 59.89 & 15.23 & 6.37 & 52.10 & 39.46 \\
GPT-5         & 79.15 & \textbf{69.79} & 17.13 & 6.87 & 56.36 & 48.12 \\
DeepSeek-Chat & 65.65 & 62.93 & 11.83 & 4.65 & 54.03 & 46.09 \\

\midrule
\rev{Claude 3.7}    & \rev{58.03} & \rev{56.51} & \rev{29.74} & \rev{29.17} & \rev{53.5} & \rev{52.43} \\
\rev{GPT-5}         & \rev{63.40} & \rev{\textbf{65.59}} & \rev{27.53} & \rev{26.33} & \rev{44.75} & \rev{43.86} \\
\rev{DeepSeek-Chat} & \rev{62.36} & \rev{61.72} & \rev{29.10} & \rev{26.96} & \rev{55.35} & \rev{55.24} \\
\midrule
\midrule
\rev{Claude 3.7}    & \rev{56.75} & \rev{48.20} & \rev{16.76} & \rev{8.05} & \rev{38.73} & \rev{26.91} \\
\rev{GPT-5}         & \rev{64.21} & \rev{\textbf{56.07}} & \rev{17.20} & \rev{8.35} & \rev{41.21} & \rev{32.20} \\
\rev{DeepSeek-Chat} & \rev{59.29} & \rev{51.70} & \rev{15.67} & \rev{6.61} & \rev{40.15} & \rev{30.72} \\
\bottomrule
\end{tabular}

\end{tabular}
\vspace{-4mm}
\end{table}

\subsection{RQ2: \projName's Applicability To Real-World Projects}

We evaluate \projName{} on FDLibM and the structured-input programs separately, since they involve different baselines.
Table~\ref{tab:FDLibM-summary} reports line coverage on FDLibM, where coverage reflects each technique's ability to explore floating-point-guarded control flow rather than reach explicit bomb branches. KLEE achieves 56.7--63.1\% average and up to 70.2\% total coverage; STP and Z3 perform comparably, with STP marginally ahead due to higher solver throughput. KLEE-Float-Z3 trails despite more precise floating-point reasoning, for two reasons: slow float solving limits exploration (only 21.37\% as many test cases as standard KLEE), and its 9-year-old KLEE fork lacks modern performance optimizations.

ConcoLLMic, in contrast to its LogicBombs results, performs poorly on FDLibM (37.2--40.0\% average, 29.0--31.1\% total coverage). Even its best run with Claude-3.7 yields 55.66\% less coverage than KLEE-STP despite issuing up to 16.45M tokens. This exposes a fundamental weakness of whole-path LLM reasoning on numerical code: predicting inputs that satisfy long chains of floating-point constraints is unreliable from semantic reasoning alone.

\projName{} achieves the highest coverage on FDLibM, up to 91.7\% average and 90.2\% total---an average 183.5\% improvement over ConcoLLMic and 22.09\% over the best KLEE version, despite generating only 13.87\% as many test cases. To rule out better time-budget allocation as the explanation, we combine the coverage of all three KLEE variants and all three ConcoLLMic configurations, a baseline with 6$\times$ the exploration time of a single \projName{} run. This union reaches only 70.75\% average and 77.95\% total coverage, while each individual \projName{} configuration exceeds it by 6.88--15.75\% in total coverage. \projName{}'s gains thus come from reaching solver-hostile branches that remain inaccessible even when the strengths of existing approaches are aggregated.





Table~\ref{tab:coverage-summary} reports line coverage on the structured-input benchmarks, which exercise grammar-constrained parsing and semantic validation. KLEE variants achieve modest coverage, with total coverage below 40\% even for the best configuration. Z3 and STP perform comparably, while Float-Z3 trails sharply---most notably on \texttt{bc}, where missing inline-assembly support limits it to a single test case---and elsewhere generates under 1\% as many tests as standard KLEE.

Neither ConcoLLMic nor Cottontail consistently outperforms KLEE on structured-input programs. ConcoLLMic falls below 10\% total coverage---only \rev{29.44\%} of KLEE-STP's coverage on average---largely due to instrumentation failures: functions exceeding LLM output limits, and instrumentation points that cannot safely accommodate injected logging, both yielding zero coverage for the affected files. Cottontail fares better since its grammar-based generation matches structured-input programs naturally, beating the best KLEE variant by \rev{23.61\%} with GPT-5; it is strongest on \texttt{libyaml} and \texttt{bc} but struggles on \texttt{libexpat}, where coverage requires deep semantic validation beyond syntactic correctness. Overall, end-to-end LLM-driven concolic execution remains fragile in the presence of complex cross-function dependencies.

\projName{} outperforms all baselines on every structured-input benchmark and every backend, improving total coverage by \rev{85.03--115.24\%} over the strongest KLEE, by \rev{6.8$\times$} on average over ConcoLLMic, and by \rev{83.36\% on average over Cottontail (29.94\% $\to$ 51.9\% total coverage)}.

\begin{tcolorbox}[colback=gray!5, colframe=black!20, arc=2pt, boxrule=0.3mm, breakable, sharp corners, left=2pt, right=2pt, top=2pt, bottom=2pt]
\textbf{RQ2:} \projName{} scales  to real-world numerical and structured-input programs, consistently outperforming symbolic execution and LLM-based baselines.
On FDLibM, \projName{} improves coverage by \textbf{183.5\%} on average over ConcoLLMic and by \textbf{22.09\%} over the strongest KLEE configuration.
On structured-input programs, \projName{} improves total coverage by up to \textbf{\rev{115.24\%}} over KLEE, by \textbf{\rev{83.36\%}} on average over Cottontail, and covers \textbf{\rev{6.8}$\times$} more lines than ConcoLLMic.
\end{tcolorbox}

\begin{revblock}

\begin{table}[t]
\centering
\begin{minipage}[b]{0.48\linewidth}
\centering

{\footnotesize
\begin{tabular}{lr}
\hline
\rev{Outcome category} & \rev{Attempts (\%)} \\
\hline
\rev{A1-Generation/execution failure} & \rev{7} \\
\rev{A2-Interface mismatch } & \rev{8} \\
\rev{A3-Semantically invalid inversion} & \rev{22} \\
\rev{P1-Non convergence} & \rev{18} \\
\rev{P2-Successful path trigger} & \rev{45} \\
\hline
\end{tabular}
}
\caption{\rev{Mutually exclusive categories of the outcomes for 100 inversion attempts from LogicBombs.}}
\label{tab:inversion_pipeline}
\end{minipage}
\hfill
\begin{minipage}[b]{0.48\linewidth}
\centering
{\footnotesize
\begin{tabular}{lcc}
\hline
\rev{Type} & \rev{\makecell{Additional \\ Coverage (\%)}} & \rev{\makecell{Avg. \\ Attempts}} \\
\hline
\rev{Partition} & \rev{93.2} & \rev{1.2} \\
\rev{Modeling} & \rev{62.1} & \rev{1.4} \\
\rev{Inversion} & \rev{43.2} & \rev{2.3} \\
\hline
\end{tabular}
}
\caption{\rev{Effectiveness of ghost-code strategies across LogicBombs, fdlibm, structured inputs benchmarks.}}
\label{tab:strategy_effectiveness}
\end{minipage}
\label{tab:inversion_results}
\vspace{-8mm}
\end{table}

We analyze 100 randomly sampled inversion attempts from LogicBombs, excluding malformed outputs such as truncated or unparsable responses, classifying their first failure stage (Table~\ref{tab:inversion_pipeline}). The first three categories are \emph{artifact defects}; the last two are at least approximations of the true inverse.

\begin{description}\itemsep1pt
\item[(A1) Generation/execution failure (7\%).] Does not compile, crashes, or cannot run.
\item[(A2) Interface mismatch (8\%).] Runs but output is unusable by Gordian's refinement (missing variables, wrong type, schema violation).
\item[(A3) Semantically invalid (22\%).] Correct interface, but not a valid inverse or approximation.
\item[(P1) Non-convergence (18\%).] Refinement does not converge within budget. Manual inspection suggests the candidates are plausible approximations; the loop fails to close mainly due to infeasible target paths or approximation error too large for the optimization step to absorb, not inversion defects. This is the operational \textsc{Unsat} case of \Cref{sec:bidirectional}.
\item[(P2) Successful path trigger (45\%).] The inversion yields a concrete input validated on the original program that reaches the uncovered path.
\end{description}

Two rates summarize the distribution. The artifact-level success rate (defect-free inversions, A1--A3 excluded) is $63\%$, with $93\%$ compiling and $85\%$ executing. The end-to-end rate is $45\%$; the $18\%$ gap is category P1, where the inversion is plausible but the refinement loop does not close---a pipeline property, not an inversion defect. Many successful inversions are themselves approximate (search-based or heuristic) rather than exact, yet propagate values consistent enough for refinement to converge. LLMs are thus reliable enough as inversion generators to aid symbolic execution.

Across FDLibM and the structured-input benchmarks, \projName{} identifies one solver-hostile fragment per \(117.5\) executable lines on average, targeted with one or more of the three ghost-code strategies (Table~\ref{tab:strategy_effectiveness}). Partitioning is the most effective, achieving the highest coverage with minimal attempts. Modeling balances generality and effectiveness. Inversion is less consistent and needs more attempts, but is crucial in solver-hostile numerical settings where other strategies fail.
\end{revblock}

\subsection{RQ3: \projName's Token Efficiency}


\projName{} uses the LLM fundamentally differently from ConcoLLMic and Cottontail: it invokes the LLM only at specific solver-hostile fragments to generate a small amount of ghost code, then delegates the rest to the symbolic executor (\rev{LLM calls account for only ${\sim}9.2\%$ of Gordian's execution time}). ConcoLLMic and Cottontail, by contrast, query the LLM continuously to reason about execution paths and synthesize inputs, incurring LLM cost throughout exploration.

\begin{wraptable}{r}{0.5\textwidth}
  \vspace{-3mm}
\centering
\caption{Total token usage (in Millions) across structured-input benchmarks.}
\label{tab:token-usage}
\footnotesize
\begin{tabular}{lccc}
\toprule
\rev{Model} & \rev{\projName{}} & \rev{ConcoLLMic} & \rev{Cottontail} \\
\midrule
\rev{Claude 3.7}    & \rev{0.40} & \rev{24.41} & \rev{6.87} \\
\rev{GPT-5}         & \rev{0.47} & \rev{14.90} & \rev{2.81} \\
\rev{DeepSeek-Chat} & \rev{0.40} & \rev{14.91} & \rev{5.02} \\
\bottomrule
\end{tabular}
  \vspace{-2mm}
\end{wraptable}

On LogicBombs (\Cref{tab:logic-bombs-summary}), \projName{} uses roughly $10\times$ fewer tokens than ConcoLLMic (0.32M vs.\ 3.07M per run on average). On FDLibM (\Cref{tab:FDLibM-summary}), the reduction grows to $23\times$. On the structured-input benchmarks (Table~\ref{tab:token-usage}), ConcoLLMic consumes on average \rev{42.69}$\times$ more tokens than \projName{}, while Cottontail consumes between \rev{5.98}$\times$ (GPT-5) and \rev{17.17}$\times$ (Claude-3.7). \projName{} thus achieves higher coverage at substantially lower LLM cost than existing LLM-based testing approaches.

\begin{tcolorbox}[colback=gray!5, colframe=black!20, arc=2pt, boxrule=0.3mm,
                  breakable, sharp corners, left=2pt, right=2pt, top=2pt, bottom=2pt]
\textbf{RQ3:} \projName{} achieves more than an order-of-magnitude reduction
in LLM token cost while achieving higher coverage.
Across all 5 benchmarks, ConcoLLMic consumes \textbf{\rev{26.38}$\times$} as many tokens
as \projName{} (a \textbf{\rev{96.21}\%} reduction), while
Cottontail consumes \textbf{\rev{11.57}$\times$} as many tokens
(a \textbf{\rev{91.36}\%} reduction for \projName{}).
\end{tcolorbox}

\begin{revblock}
\subsection{RQ4: \projName{}'s Coverage Progress Over Time}

To assess whether \projName{}'s gains are realized only after long runs, we measure coverage as a function of wall-clock time across all evaluated tools.

\begin{figure}[t]
    \centering
    \includegraphics[width=0.49\textwidth]{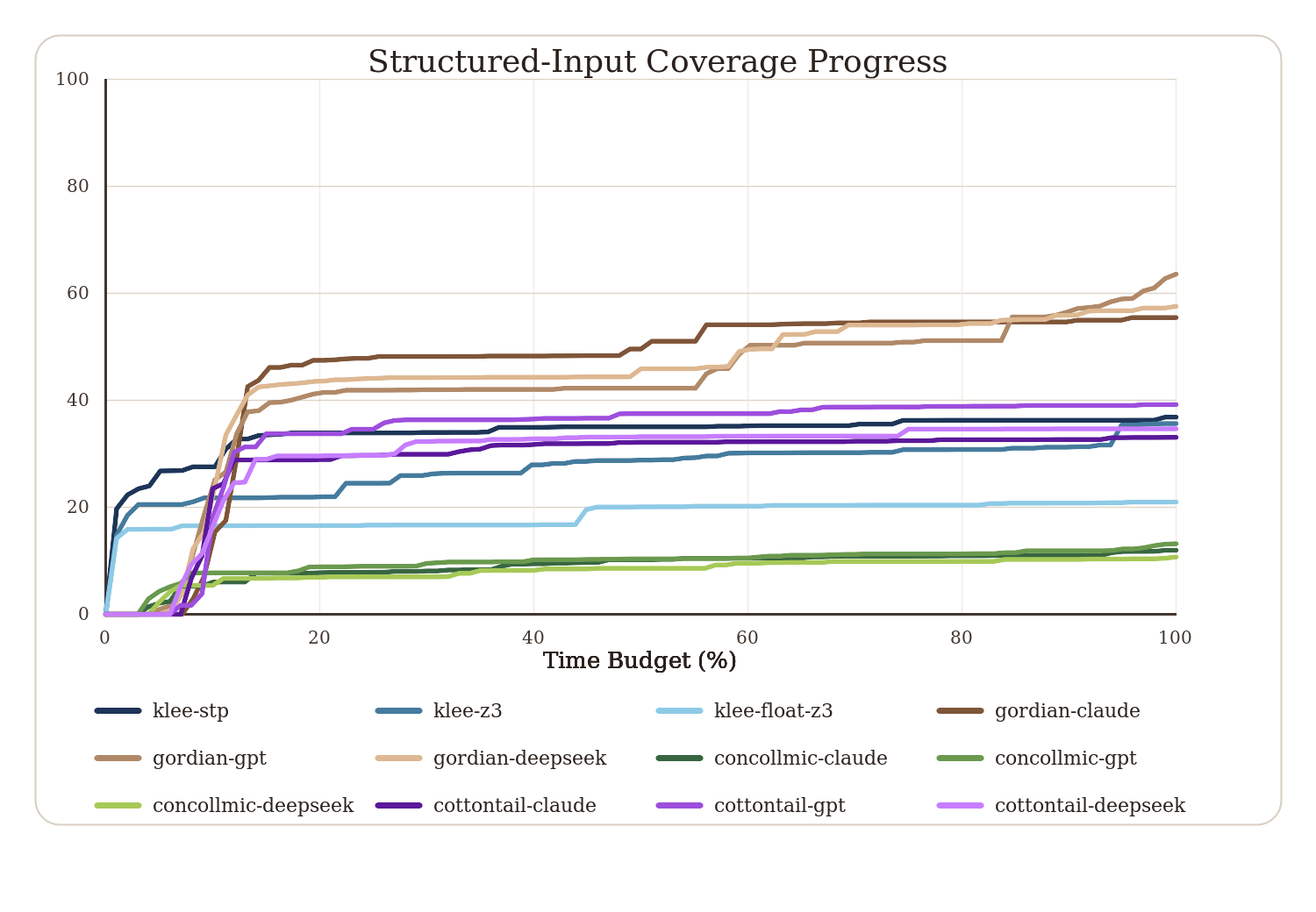}
    \hfill
    \includegraphics[width=0.49\textwidth]{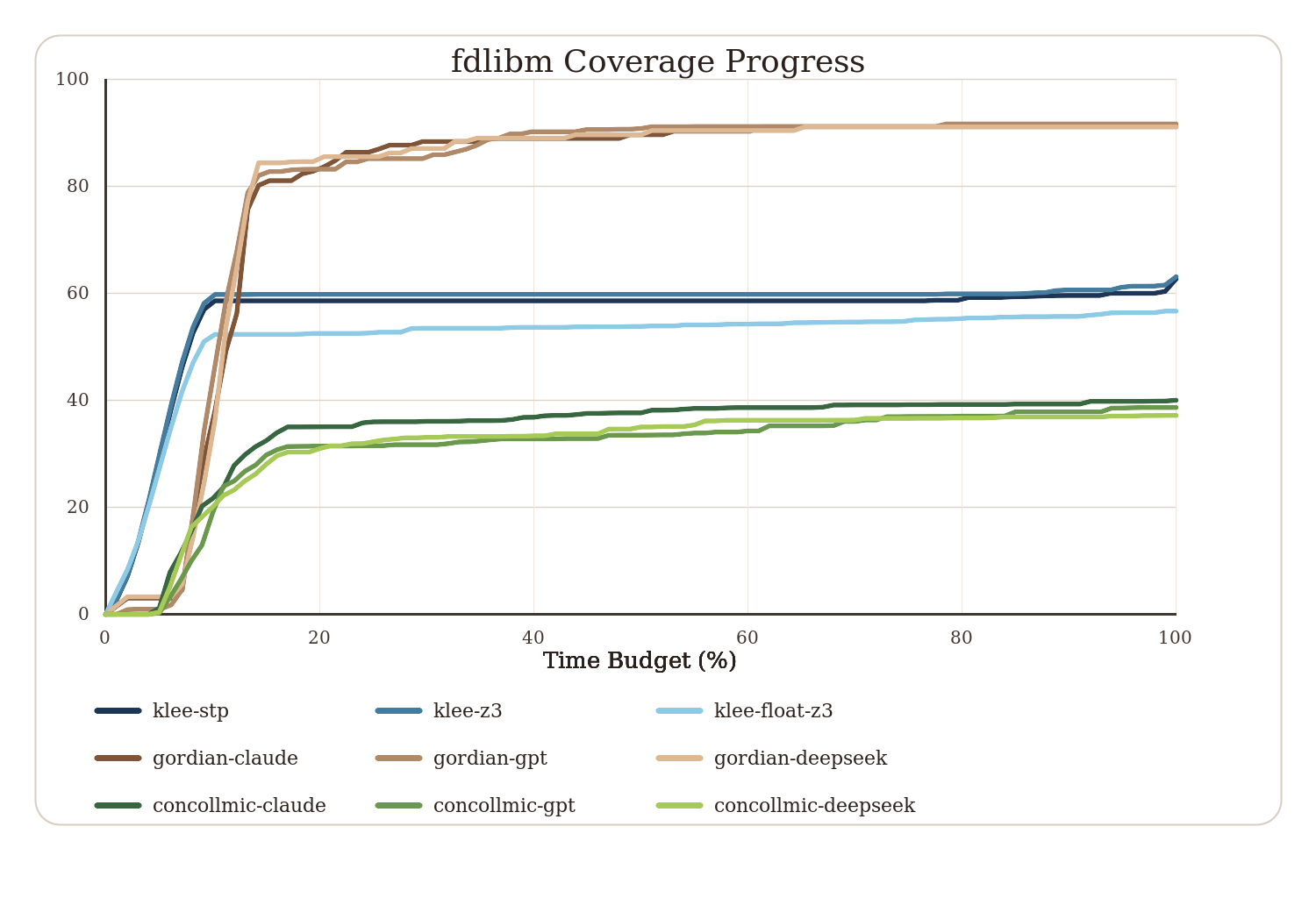}
    \vspace{-7mm}
    \caption{\begin{revblock}
        Coverage progression over time for structured-input programs (left) and FDLibM (right).
    \end{revblock}}
    \label{fig:coverage-curves}
    \vspace{-4mm}
\end{figure}

Figure~\ref{fig:coverage-curves} shows coverage versus time budget. All approaches follow a common pattern---rapid initial gains followed by a plateau---but differ in where their gains concentrate. KLEE begins exploring immediately and accrues coverage early, while LLM-based approaches incur an initial delay for preprocessing and LLM calls; once exploration starts, their growth trends resemble KLEE's.

\projName{}'s curves are distinctly step-wise: large early jumps correspond to successful ghost-code transformations (partitioning, modeling) that unlock previously unreachable regions, and subsequent improvements occur in discrete steps as new partitions or solver-aiding transformations are introduced. \projName{} typically surpasses both KLEE and ConcoLLMic well within the first few hours, and the bulk of its advantage over KLEE is realized before the long-budget plateau---meaning the gains are not an artifact of the 36-hour ceiling. Cottontail and ConcoLLMic, by contrast, grow more smoothly because their LLM interactions are interleaved continuously with exploration.

On FDLibM, KLEE quickly generates floating-point-heavy test cases for high initial coverage, but stagnates early as many paths hit solver-hostile constraints it cannot extend. \projName{} continues to make progress by resolving such bottlenecks via targeted transformations. Minor spikes in the KLEE curves are a measurement artifact: coverage updates only when test cases are generated, and several paths sharing prefixes can reach new code within a short window.

\begin{tcolorbox}[colback=gray!5, colframe=black!20, arc=2pt, boxrule=0.3mm, breakable, sharp corners, left=2pt, right=2pt, top=2pt, bottom=2pt]
\rev{\textbf{RQ4:} \projName{}'s advantage emerges early rather than only at long budgets. Many of its gains over KLEE and ConcoLLMic appear within the first 20\% of the 36-hour budget, while additional gains continue later through step-wise improvements unlocked by ghost-code transformations.}
\end{tcolorbox}
\end{revblock}

\begin{revblock}
    
\subsection{RQ5: \projName{}'s Exploration of Hard-to-Reach Paths}
\label{sec:hard-paths}

To check whether \projName{}'s gains correspond to genuinely new paths rather than redundant coverage, we analyze the overlap of covered lines across techniques. Coverage is a standard testing-effectiveness proxy in symbolic and concolic execution and provides more stable evaluation signals than bug-based metrics~\cite{FSE26-concordance}; the overlap view further reveals how much coverage is shared and how much each technique contributes uniquely.

\begin{figure}[t]
\centering
\includegraphics[width=0.9\textwidth]{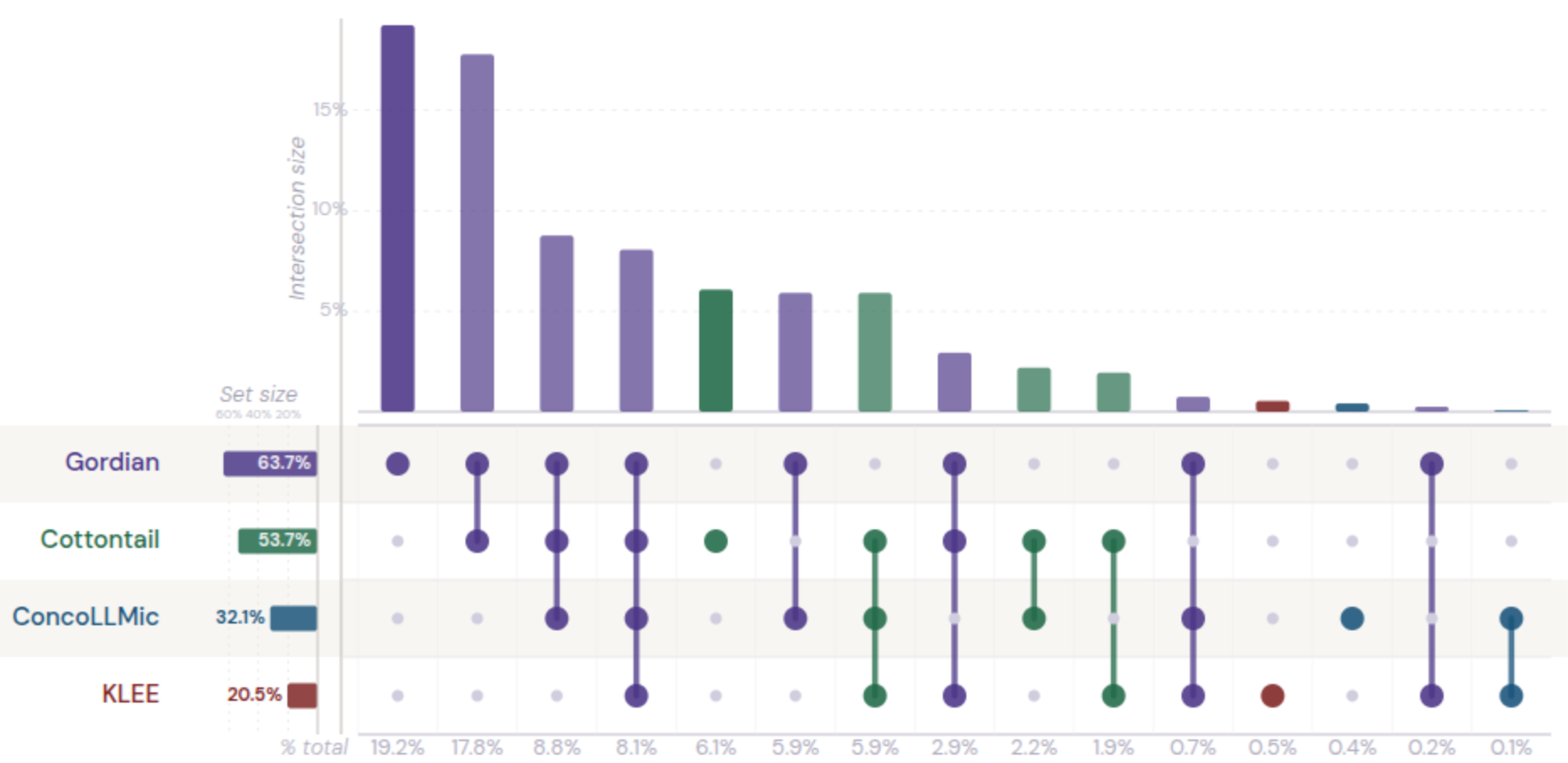}

\caption{\begin{revblock}
Overlap of covered lines across techniques on \texttt{libyaml}, aggregated across all models.
The largest intersection corresponds to lines covered exclusively by \projName{} (19.2\% of all covered lines), exceeding even the shared \projName{}--Cottontail intersection (17.8\%).
\end{revblock}}
\label{fig:libyaml-upset}
\end{figure}

Figure~\ref{fig:libyaml-upset} shows the overlap on \texttt{libyaml}, aggregated across LLM backends (the best-performing structured-input benchmark for LLM-based approaches). Three observations stand out: \projName{} achieves the largest overall coverage; it contributes a substantial set of lines covered by no other approach, reaching previously unexplored behaviors; and although it overlaps significantly with structure-aware Cottontail, it still extends coverage beyond their union---its gains are not reducible to input-structure modeling alone.

\begin{tcolorbox}[colback=gray!5, colframe=black!20, arc=2pt, boxrule=0.3mm, breakable, sharp corners, left=2pt, right=2pt, top=2pt, bottom=2pt]
\rev{\textbf{RQ5:} \projName{} reaches hard coverage regions not attained by the other evaluated techniques. On \texttt{libyaml}, \textbf{19.2\%} of covered lines are reached by \projName{} alone---exceeding even its overlap with Cottontail---so its advantage cannot be explained by input-structure modeling or redundant coverage of paths reachable by other techniques.}
\end{tcolorbox}
\end{revblock}

\begin{revblock}
  \subsection{Rerun Variability Analysis.}
  
To quantify the impact of LLM stochasticity, we evaluate a representative subset of 16 source files (around 10\% of the total), uniformly sampled across benchmarks. For each file and each of the three LLM backends, we perform three independent reruns, yielding 48 benchmark–model pairs and 144 executions in total. 

We measure variability using the within-pair standard deviation across reruns. The average within-pair standard deviation is 3.08 coverage points overall, with per-model values of 2.65 (Claude), 2.36 (GPT), and 4.22 (DeepSeek), indicating moderate stochasticity at the level of individual runs.

To assess the impact on results, we approximate the standard error of average coverage over $N$ files as $\mathrm{SE} \approx s / \sqrt{N}$. This yields an approximate 95\% stochasticity band of ±0.68 coverage points for FDLibM (78 files; ±0.59 for Claude, ±0.52 for GPT, and ±0.94 for DeepSeek) and ±1.38 points for structured-input programs (19 files; ±1.19 for Claude, ±1.06 for GPT, and ±1.90 for DeepSeek). Per-model bounds remain below ±1 point for FDLibM and below ±2 points for structured benchmarks.
These deviations are small relative to the observed improvements (double digits coverage points over baselines), indicating that our findings are robust to LLM stochasticity.
\end{revblock}

\section{Discussion}

Although \projName significantly outperformed baselines in the evaluation, it has several limitations. First, \projName operates by injecting LLM-generated ghost code and recompiling the target program for symbolic execution. \rev{This introduces engineering overhead, especially for large systems with complex build pipelines, external dependencies, or custom build scripts. In practice, the overhead is mitigated by incremental builds: \projName{} modifies only the harness or a small number of target/source files, so build systems such as \texttt{make} typically recompile only the affected objects and relink the executable rather than rebuilding the entire project. Nevertheless, manual integration effort may still be required for complex projects.} Second, \projName relies on general-purpose LLMs, which may reflect biases or implicit knowledge from their training data. In addition, LLM-generated ghost code may be incomplete or imprecise, particularly for inversion and semantic modeling. Importantly, \projName{} treats all LLM outputs as heuristic aids: all generated test inputs are replayed on the original program, ensuring that incorrect abstractions affect efficiency rather than soundness; however, it may still miss feasible paths. Third, \projName is most effective when solver-hostile behavior is localized to identifiable program fragments. Programs whose complexity arises from deeply entangled global invariants or hard-to-model environment interactions may benefit less from ghost code generation. Fourth, \projName's bidirectional constraint propagation algorithm in \Cref{sec:bidirectional} relies on heuristic optimization, and does not guarantee convergence.

A natural direction for future work is a tighter integration of \projName's capabilities directly inside a symbolic execution engine. In our current implementation, \projName{} operates as an external orchestrator that interacts with KLEE through repeated recompilation and execution. While effective, this architecture introduces engineering complexity and multiple points of failure. A more principled approach would embed LLM assistance natively within KLEE. Solver queries or path fragments that exceed resource limits could be detected online and selectively delegated to LLM-based reasoning modules. Once a generated model or inverse is validated by successfully triggering a previously unreachable path, it could be cached and reused across executions or program versions. Over time, this would yield an evolving library of solver-friendly models and inverses, potentially compositional and shared across analyses.

\begin{revblock}

A limitation of the bidirectional constraint propagation algorithm presented in \Cref{sec:bidirectional} is that it assumes a single difficult fragment along any explored path. Accordingly, our implementation never attempts to invert more than one fragment at a time. This is a pragmatic choice rather than a fundamental restriction: empirically, addressing even a single difficult fragment yields substantial practical gains. In principle, however, the algorithm extends naturally to paths containing multiple difficult fragments, as follows. Suppose symbolic execution of the transformed program $P'$ encounters a sequence of $n$ difficult fragments $C_1,\ldots,C_n$ along a single path, each equipped with an LLM-generated inverse $f_i^{-1}$. The accumulated path condition then decomposes into $n+1$ contiguous segments,
\[
\pi \;=\; \pi_0 \,\wedge\, \pi_1 \,\wedge\, \cdots \,\wedge\, \pi_n,
\]
where $\pi_0$ constrains only the original input variables (the prefix preceding $C_1$), and each subsequent $\pi_i$ ($1\le i\le n$) is the segment between $C_i$ and $C_{i+1}$ (with $\pi_n$ the final suffix), depending on the fresh symbolic variables used in $C_1,\ldots,C_i$ as well as on the original inputs.

The generalized procedure might proceed by ``meeting in the middle'' across all $n$ cuts simultaneously. It first solves the bottom segment $\pi_n$ to obtain an assignment $A_n$, then propagates values upward fragment by fragment: from $A_i$ it extracts the values written by $C_i$, applies $f_i^{-1}$ to obtain a suggested pre-state $\sigma_i$ over the variables read in $C_i$, and checks whether $\sigma_i$ together with $A_i$ is consistent with the segment $\pi_{i-1}$. If every $\pi_{i-1}$ is satisfied along the way and reaches $\pi_0$ consistently, a model for the whole path has been found and the algorithm terminates. Otherwise, at the highest segment $\pi_{j-1}$ that becomes unsatisfied, the algorithm invokes the optimizing solver to compute an assignment $A_{j-1}\models\pi_{j-1}$ that is as close as possible to $\sigma_j$, and then propagates $A_{j-1}$ downward by concretely executing $f_{j-1}, f_j, \ldots, f_n$ in turn. The upward--downward traversal is then repeated from the new assignment, until either a fixpoint that satisfies $\pi_0\wedge\cdots\wedge\pi_n$ is reached, or the procedure stalls and conservatively returns \textsc{Unsat}. We leave the implementation, convergence analysis, and empirical evaluation of this multi-fragment extension to future work.

\end{revblock}


\section{Related Work}

\paragraph*{Symbolic and Concolic Testing} Symbolic execution~\cite{king1976symbolic} (e.g., KLEE~\cite{klee}) explores program paths over symbolic inputs, but is limited by path explosion and solver bottlenecks, especially with complex arithmetic and hard-to-model environments~\cite{env_modeling,autobug}. Concolic execution~\cite{dart} improves practicality by interleaving concrete runs with symbolic reasoning; systems such as S2E~\cite{s2e} scale this to whole-system settings, while compilation-based engines (e.g., SymCC~\cite{symcc}) improve throughput via native execution and lightweight instrumentation. Prior work addresses exploration blowups from loops and recursion (e.g., MARCO~\cite{hu2024marco}), whereas \projName targets the complementary bottleneck: solver-hostile fragments that block progress. \rev{Environment modeling via synthesis (e.g., SE-ESOC~\cite{mechtaev2018symbolic}) replaces missing or external semantics with generated code; \projName differs in that it targets internal program fragments whose code is present but solver-hostile, and uses ghost code not merely as a stub but as a solver aid for propagating and reconciling constraints across the original fragment. \projName generalizes this idea by using LLMs to generate ghost code that (i) propagates constraints through hard fragments, (ii) introduces solver-friendly surrogates that preserve relevant behavior, and (iii) constrains pointer-rich state spaces using semantic heap topologies. }For dynamic data structures, classic lazy initialization~\cite{visser2004test} incrementally materializes heap objects but remains largely shape-agnostic; \projName instead injects domain-informed topological partitions to reduce search space explosion while keeping data fields symbolic. 
\rev{\paragraph*{Hybrid Fuzzing and Symbolic Execution}} Compared to symbolic/concolic execution, fuzzing (e.g., AFL++~\cite{aflpp}) offers high throughput but typically struggles with deep, constraint-heavy branches~\cite{manes_fuzzing_survey}. \rev{Systems such as Driller~\cite{driller} and QSYM~\cite{qsym} are complementary to \projName{}: they combine fuzzing with concolic execution to improve input-space exploration, whereas \projName{} combines LLM guidance with symbolic execution to target solver-hostile internal fragments through selective solver-aiding ghost code rather than repeated input mutation or full-path reasoning.}
\paragraph*{LLM-based Symbolic and Concolic Execution} Recent work integrates machine learning and large language models (LLMs) into symbolic execution to mitigate scalability limits caused by complex constraints and environment modeling.
Early approaches focused on learned path exploration heuristics, using techniques such as Monte Carlo Tree Search~\cite{legion} or learned branch prioritization policies (Learch~\cite{learch}). More recent systems embed LLMs directly into symbolic or concolic execution. Solver-free approaches such as PALM~\cite{palm} and AutoBug~\cite{autobug} rely on LLMs to generate satisfying inputs for program fragments or path-wise slices, bypassing SMT solvers at the cost of global constraint consistency and scalability on deep paths. Cottontail~\cite{cottontail} targets structured-input programs using LLM-guided grammar-based input generation, while ConcoLLMic~\cite{concollmic} adopts an agentic design in which LLMs summarize traces and reason about constraints. Although LLMs improve coverage by bypassing solver bottlenecks, we showed that these systems are ineffective in handling complex program dependencies. Our hybrid approach that combines LLMs with SMT solvers addresses this problem.
\paragraph*{LLM-assisted Fuzzing and Test Generation}
LLMs have also been applied to fuzzing and test generation primarily as input generators, rather than as engines for symbolic reasoning. CodaMOSA~\cite{codamosa} uses language models to synthesize new tests when coverage plateaus, while other systems apply LLM-driven mutations to protocol and application fuzzing~\cite{chatfuzz,inputblaster}. Although effective at expanding the input space and improving shallow coverage, these approaches do not reason over execution paths or symbolic constraints and thus remain limited in reaching deep program states~\cite{manes_fuzzing_survey}.
\paragraph*{Program Inversion}
Program inversion aims to generate an inverse program that computes inputs that lead to a given output. Previous work synthesized inverse computations using symbolic execution and solvers~\cite{srivastava2011path,Abramov2000InverseComputation}, however these techniques are typically limited to small fragments or solver-friendly semantics, making them unsuitable for our use case. Just-tri-it~\cite{dai2025reducing} uses LLM-generated algorithm inverses to check cross-task solution consistency, which improves the quality of generated code. \projName applies an LLM to invert only a small program fragment, which is easier, and uses such inverses to propagate values across execution paths. \rev{Unlike prior inversion work, \projName does not require synthesizing a complete inverse program; it uses approximate fragment inverses inside a bidirectional meet-in-the-middle loop that combines LLM-generated code with SMT reasoning.}
\paragraph*{Reasoning About Programs with LLMs} Recent work has studied and improved LLMs' ability to reason about code~\cite{chen2024reasoning,liu2025evaluating}. HoarePrompt~\cite{bouras2025hoareprompt} structures chain-of-thought using ideas from strongest postcondition calculus, closely related to symbolic execution. AutoBug~\cite{autobug} improves LLM reasoning by decomposing programs into simpler components. \projName is complementary to these approaches: advances in LLM reasoning can directly improve the quality of the ghost code.


\section{Conclusion}

This work presents \projName, a hybrid symbolic execution framework that addresses key bottlenecks of SMT-backed symbolic execution using LLMs without giving up the ability to enforce globally consistent, cross-procedural constraints. Rather than replacing the solver with an LLM, \projName uses the LLM selectively to generate lightweight, solver-aiding ghost code tailored to solver-hostile fragments: inverse procedures for bidirectional constraint propagation, solver-friendly surrogate models that preserve control-relevant behavior, and semantic heap topologies that curb path explosion in pointer-rich code. Implemented on top of KLEE and evaluated across synthetic logic bombs, a real mathematical library, and large structured-input applications, \projName consistently improves coverage and reduces LLM cost relative to both traditional symbolic execution and prior LLM-driven alternatives.

\section*{Data Availability}

All experimental artifacts required to reproduce the results of in this paper are publicly available at:
\url{https://figshare.com/s/09550df85983cbf4ae71}. 
The artifact package includes: (i) the complete implementation of \emph{\projName{}}, (ii) all baseline implementations and
  artifacts for KLEE, ConcoLLMic, and Cottontail, (iii) benchmark programs and harnesses for all six evaluated benchmark
  groups (LogicBombs, FDLibM, GNU \texttt{bc}, \texttt{jq}, \texttt{libexpat}, and \texttt{libyaml}), (iv) generated test
  cases, execution logs, and intermediate artifacts, and (v) post-processed results used to produce the tables and newly added
  analyses in the paper.

\section*{Acknowledgments}
This work was supported by the National Natural Science Foundation of China (NSFC) under Grant No.~W2542035 and Grant No.~92582202. We thank Bo Wang for insightful discussions.

\bibliography{refs}
\bibliographystyle{plain}

\end{document}